\documentclass[aps,prb,twocolumn,amsmath,amssymb,floatfix]{revtex4-2}
\usepackage[english]{babel}
\usepackage{blindtext}
\usepackage[utf8]{inputenc}
\usepackage{graphicx}
\usepackage{array} 
\usepackage{bm}
\usepackage{latexsym}
\usepackage{physics}
\usepackage{amsfonts}
\usepackage{bbold}
\usepackage{subfig}
\usepackage{scalerel}
\usepackage{appendix}
\usepackage{physics,braket}
\usepackage{ragged2e}
\usepackage{mathtools}

\usepackage[colorlinks = true,
            linkcolor = blue,
            urlcolor  = blue,
            citecolor = blue,
            anchorcolor = blue]{hyperref}
\usepackage[usenames,dvipsnames]{color}
\makeatletter
\newcommand*{\rom}[1]{\expandafter\@slowromancap\romannumeral #1@}
\newcommand{\vebm}[1]{
\ifcat\noexpand#1\relax
    {\bm #1}
\else
    {\bf #1}
\fi
}
\newsavebox{\@brx}
\newcommand{\llangle}[1][]{\savebox{\@brx}{\(\m@th{#1\langle}\)}%
  \mathopen{\copy\@brx\kern-0.5\wd\@brx\usebox{\@brx}}}
\newcommand{\rrangle}[1][]{\savebox{\@brx}{\(\m@th{#1\rangle}\)}%
  \mathclose{\copy\@brx\kern-0.5\wd\@brx\usebox{\@brx}}}
\makeatother
\begin{document}

\title{Topological responses from gapped Weyl points in 2D altermagnets}

\date{\today}

\author{Kirill Parshukov}
\email{k.parshukov@fkf.mpg.de}
\thanks{Shared first author.}
\affiliation{Max Planck Institute for Solid State Research, Heisenbergstrasse 1, D-70569 Stuttgart, Germany}

\author{Raymond Wiedmann}
\email{r.wiedmann@fkf.mpg.de}
\thanks{Shared first author.}
\affiliation{Max Planck Institute for Solid State Research, Heisenbergstrasse 1, D-70569 Stuttgart, Germany}

\author{Andreas P. Schnyder}
\email{a.schnyder@fkf.mpg.de}
\affiliation{Max Planck Institute for Solid State Research, Heisenbergstrasse 1, D-70569 Stuttgart, Germany}

\begin{abstract}

We study the symmetry requirements for topologically protected spin-polarized Weyl points in 2D altermagnets. The topology is characterized by a quantized $\pi$-Berry phase and the degeneracy is protected by spin-space group symmetries. Gapped phases with finite Chern and/or spin/chirality Chern numbers emerge under different symmetry-breaking mass terms. We investigate the surface and transport properties of these gapped phases using representative electronic tight-binding and magnonic linear spin-wave models. 
In particular, we calculate the electronic and magnonic Hall currents and discuss implications for experiments.
\end{abstract}

\maketitle
\address{$^{\ast}$ These two authors contributed equally.}
Two-dimensional (2D) materials have attracted much attention since the discovery of graphene, showcasing unique electronic, mechanical, and thermal properties~\cite{Geim2009Graphene, Mir2014AnAO}.
Graphene is the prime example of a
topological band structure in 2D,
as it exhibits  protected Dirac point crossings
between spin-degenerate bands.
A different type of topological band crossings, Weyl points (WPs), can emerge by splitting the spin degeneracy, either with strong spin-orbit coupling, an applied magnetic field, or by ferro- or ferrimagnetic order.  
Recently, a variety of 2D ferromagnetic Weyl semimetals has been found, which exhibit fully spin-split band structures~\cite{Niu_2019,Jeon2019FMWP,Shuyuan2020}. In antiferromagnetic materials, the bands are spin degenerate and WPs are generally not possible. 
However, a newly established type of collinearly ordered magnetic system, altermagnets, exhibit compensated collinear magnetic order, as in antiferromagnets, with spin-split bands, as in ferromagnets, while having negligible spin-orbit coupling~\cite{Smejkal22Altermagnetism}. 
In the last few years, there has been an increasing interest in altermagnetic materials
both from the theoretical~\cite{Smejkal22Altermagnetism,Smejkal22Landscape,smejkal_nat_review_22,PhysRevX.12.040002,mcclarty2023landau,PhysRevB.105.064430,PhysRevB.102.144441,doi:10.7566/JPSJ.88.123702,PhysRevLett.132.056701}
and the experimental side
~\cite{zhu_nature_24,krempasky_jungwirth_nature_24,fedchenko_sci_advances_24,nag2023gdalsi,lin2024observation}.
These phases are promising for spintronic applications~\cite{Smejkal22Landscape} due to large spin splitting and robustness against magnetic field perturbations, and give rise to several interesting phenomena in conjunction with superconductivity,
including 
unconventional Josephson effects~\cite{PhysRevLett.131.076003,PhysRevB.108.075425,Zhang2024},
diode effects~\cite{banerjee2024altermagnetic}, Cooper-pair splitting~\cite{giil2024quasiclassical,nagae2024spinpolarized},
and topological superconductivity~\cite{chakraborty2023zerofield,PhysRevB.108.224421,ghorashi2023altermagnetic,PhysRevB.108.205410,PhysRevB.108.184505}.

In this work, we study the symmetries allowing for spin-polarized WP degeneracies in 2D altermagnets and 
derive a classification of all spin-wallpaper groups
that protect these WPs.
We construct a representative electronic tight-binding and a magnon linear spin-wave model for one of the symmetry groups. 
Considering different symmetry-breaking terms,
a variety of different topological phases emerge, including
Chern and quantum spin-Hall insulators.
For each of these phases, we discuss topological responses in terms of anomalous electronic and thermal conductivities, including the spin Hall and spin Nernst effects~\cite{annurev-conmatphys-031620-104715,Matsumoto2011thermal,Cheng2016Nernst}. 
\begin{figure}
    \centering
    \includegraphics[width=\linewidth]{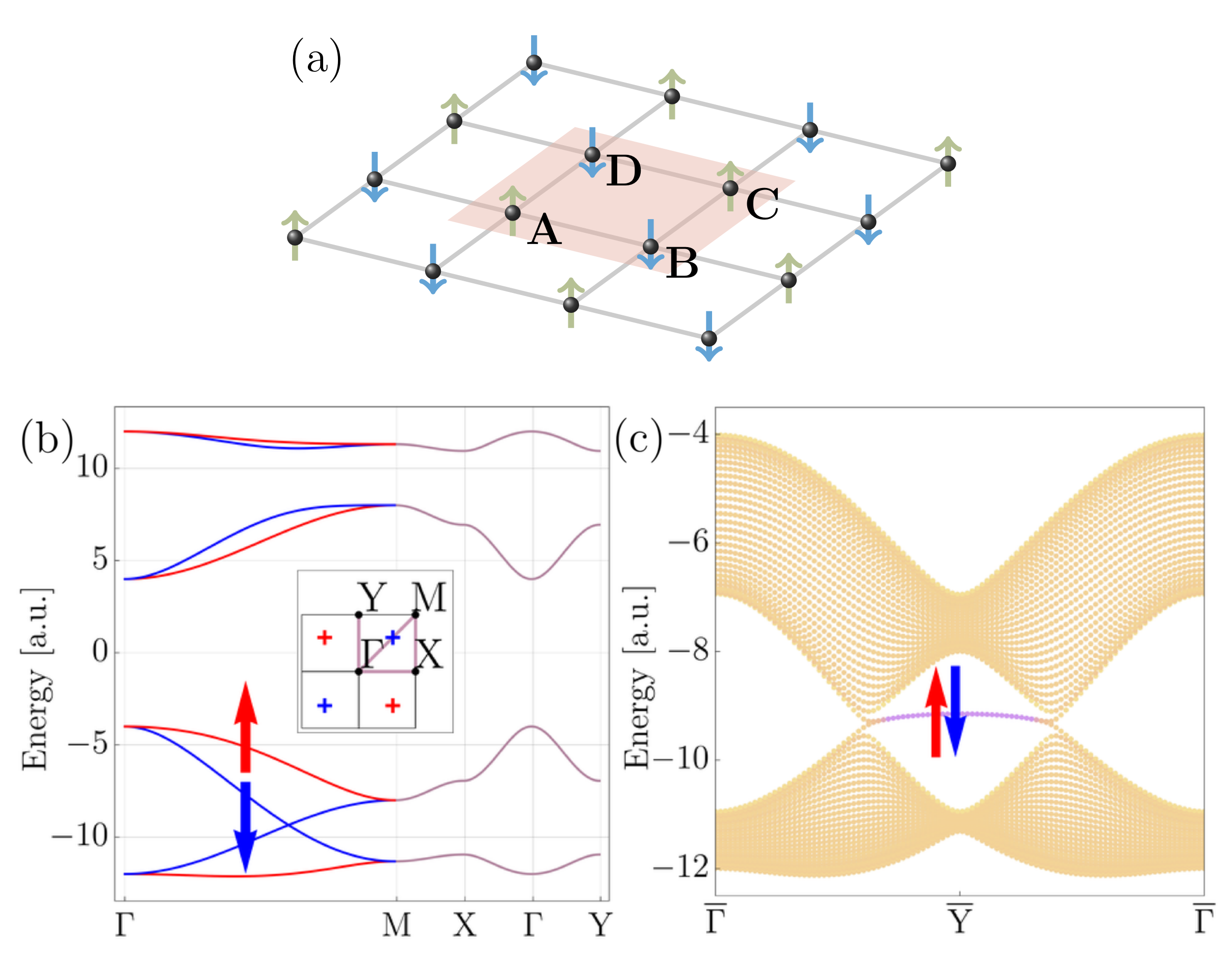}
   \caption{\justifying (a) Square lattice with collinear magnetic Néel order and four sites per unit cell (red area).
   (b) Electronic bands of the altermagnetic tight-binding Hamiltonian~\eqref{TBelectronicmodel_main} with the spin quantum number indicated by the arrows and red/blue color. 
   The inset shows the location of the
   four WPs in the 2D BZ.
  (c) Edge bandstructure
  of the electronic states within the 10 outermost unit cells of 
  a 30-cell ribbon terminated in the $x$-direction.
  }
    \label{fig:SWG6}
\end{figure}

\textit{Weyl points protected by altermagnetic symmetries.---}
The symmetries of altermagnets are described by spin-space groups~\cite{https://doi.org/10.1002/pssb.2220100206,doi:10.1098/rspa.1966.0211,10.1063/1.1708514,LITVIN1974538,xiao2023spin,ren2023enumeration,jiang2023enumeration,chen2023spin,schiff2023spin}, for which the symmetry action on the lattice and the spin degrees of freedom are decoupled. A group element is denoted by $(s||g)$, where $s$ and $g$ act on the spin and the lattice, respectively. The collinear spin arrangement along the $z$-axis allows for the SO(2) spin rotation and the $(2_x \mathcal{T}||E)$ operation. The element $(2_x \mathcal{T}||E)$ flips the spin twice with the 2-fold spin rotation $2_x$ and time-reversal operator $\mathcal{T}$ that also acts on the lattice degrees of freedom as complex conjugation. The total magnetic moment is compensated due to a symmetry that acts non-trivially on spin and lattice, such as a mirror symmetry $(2_x||\mathcal{M})$. The mirror and SO(2) spin-rotational symmetries enforce a line degeneracy along the mirror line between bands with opposite spins. 
However, a spin splitting is allowed at a general position (GP) in the Brillouin zone (BZ). 
If two bands with the same spin come close to each other, they can form a WP crossing. In 2D systems, this point degeneracy must be protected by symmetries. Generally, space-time inversion $2_z\mathcal{T}$ and mirror symmetries quantize the Berry phase protecting WPs~\cite{chiu2018quantized}. These symmetries exist in altermagnets, however, they must not act on the spin to avoid enforced spin degeneracies. 
In altermagnets with $(2_x\mathcal{T}||2_z)$ [or $(E||\mathcal{M})$] symmetry, we expect to observe protected WP degeneracies at GPs [or on mirror lines] between bands with the same spin. 
The spin polarization of the 2D WPs in altermagnets
allows for an interesting interplay of spin currents, topology, and magnetism.

In Tab.~\ref{Weyl_point_classification_new} we list all altermagnetic spin-wallpaper groups (2D spin-space groups) that can protect the WPs. Group elements that flip the spin are denoted with $\overline{1}$. Information on the enumeration can be found in the 
Supplemental Material (SM)~\cite{sm_note} and a list of the 17 wallpaper groups is found in Ref.~\cite{2016Itfc}. 
The number of WPs is always a multiple of four due to symmetries that relate the point crossings. In the remainder of this paper, we consider the altermagnetic wallpaper group p$^{\overline{1}}$4$^{\overline{1}}$m$^1$m. 
The lattice with a four-site unit cell is shown in Fig. \ref{fig:SWG6}(a). 
Mirror symmetries $(2_x||\mathcal{M}_x)$ and $ (2_x||\mathcal{M}_y)$ relate sublattices with opposite magnetic moments, i.e., under these operations site A is mapped to sites B and C, respectively. The mirrors enforce spin degeneracies along the $k_{x,y}=0,\pi$ lines. 
The spin degeneracy is lifted at GPs allowing for spin-polarized WPs protected by the $(2_x\mathcal{T}||2_z)$ symmetry. 
\begin{table}[t!]
\centering
\renewcommand{\arraystretch}{1.5} 
\begin{tabular}{ |c|cccc| } 
 \hline
 WP,  $(2_x\mathcal{T}||2_z)$ & p$^1$2$^{\overline{1}}$m$^{\overline{1}}$m, & p$^1$2$^{\overline{1}}$m$^{\overline{1}}$g, & p$^1$2$^{\overline{1}}$g$^{\overline{1}}$g, & c$^1$2$^{\overline{1}}$m$^{\overline{1}}$m,\\
 &p$^{\overline{1}}$4, & p$^{\overline{1}}$4$^{\overline{1}}$m$^1$m, & p$^{\overline{1}}$4$^1$m$^{\overline{1}}$m, & p$^1$4$^{\overline{1}}$m$^{\overline{1}}$m,\\
 &p$^{\overline{1}}$4$^{\overline{1}}$g$^1$m, & p$^{\overline{1}}$4$^1$g$^{\overline{1}}$m, & p$^1$4$^{\overline{1}}$g$^{\overline{1}}$m, &  p$^1$6$^{\overline{1}}$m$^{\overline{1}}$m \\
 \hline
 WP, $(E||\mathcal{M})$ & p$^{\overline{1}}$4$^{\overline{1}}$m$^1$m,&  p$^{\overline{1}}$4$^1$m$^{\overline{1}}$m, & p$^{\overline{1}}$4$^{\overline{1}}$g$^1$m, &  p$^{\overline{1}}$4$^1$g$^{\overline{1}}$m \\
 \hline
\end{tabular}
\caption{\justifying Altermagnetic spin-wallpaper groups that protect WPs between bands with the same spin. }\label{Weyl_point_classification_new}
\end{table}

\textit{Electronic model.---}
First, we consider an electronic tight-binding model with symmetries of the altermagnetic wallpaper group p$^{\overline{1}}$4$^{\overline{1}}$m$^1$m (for the symmetry representations see SM~\cite{sm_note}). 
The Wyckoff positions $(\tfrac{1}{4},\tfrac{1}{4}, \uparrow)$, $(-\tfrac{1}{4},\tfrac{1}{4}, \downarrow)$, $(-\tfrac{1}{4},-\tfrac{1}{4}, \uparrow)$ and $(\tfrac{1}{4},-\tfrac{1}{4}, \downarrow)$ correspond to the sites A, B, C, and D. The alignment of the magnetic moments is along the $z$-axis, so that at A, C (B, D) sites they point in the same direction. The symmetric model in $k$-space is given by
\begin{subequations}\label{TBelectronicmodel_main}
\begin{align}
    H(\mathbf{k}) &= \begin{pmatrix}
        H_0(\mathbf{k}) + N \Sigma_z&0\\
        0 & H_0(\mathbf{k}) - N \Sigma_z\\
    \end{pmatrix},\\
    H_0(\mathbf{k}) &=  \begin{pmatrix}
        \epsilon&t_{\text{AB}}&t_{\text{AC}}&t_{\text{AD}}\\
        t_{\text{AB}}^* & \epsilon&t_{\text{BC}}&t_{\text{BD}}\\
        t_{\text{AC}}^* & t_{\text{BC}}^*&\epsilon&t_{\text{CD}}\\
        t_{\text{AD}}^* & t_{\text{BD}}^*&t_{\text{CD}}^*&\epsilon\\
    \end{pmatrix},
\end{align}
\end{subequations}
where we take $\epsilon =0$, $t_{\text{AB}}=-t_{\text{CD}}=4i\sin\tfrac{k_x}{2}$, $t_{\text{AC}}=2(\cos\tfrac{k_x-k_y}{2} + \cos\tfrac{k_x+k_y}{2} - i \sin\tfrac{k_x+k_y}{2})$, $t_{\text{AD}}=t_{\text{BC}}=4i \sin\tfrac{k_y}{2}$, $t_{\text{BD}}=2(\cos\tfrac{k_x-k_y}{2} + \cos\tfrac{k_x+k_y}{2} + i \sin\tfrac{k_x-k_y}{2})$, and $N = 8$. The matrix $\Sigma_z = \text{diag} (+1, -1, +1, -1)$ describes an interaction with the staggered magnetic moments.
The band structure shown in Fig.~\ref{fig:SWG6}(b) has enforced line spin degeneracies along the $k_x$ and $k_y$ mirror lines and the bands are spin split at GPs. Four spin-polarized WPs emerge along the $\Gamma$M lines and are related by the symmetries of the model [Fig.~\ref{fig:SWG6}(b)]. Every crossing carries a $\pi$-Berry phase and has a spin flavor. 
The non-trivial Berry phase along a path through the whole BZ indicates,
by the bulk-boundary correspondence, the emergence of edge states~\cite{chiu2018quantized,PhysRevB.95.035421}. As the Hilbert space contains two decoupled subspaces with well-defined spin, the Berry phase and bulk-boundary correspondence can be introduced for them separately. Two boundary states with opposite spin emerge at each of the two edges [see Fig.~\ref{fig:SWG6}(c)].
 
\begin{figure*}[ht]
    \includegraphics[width=\textwidth]{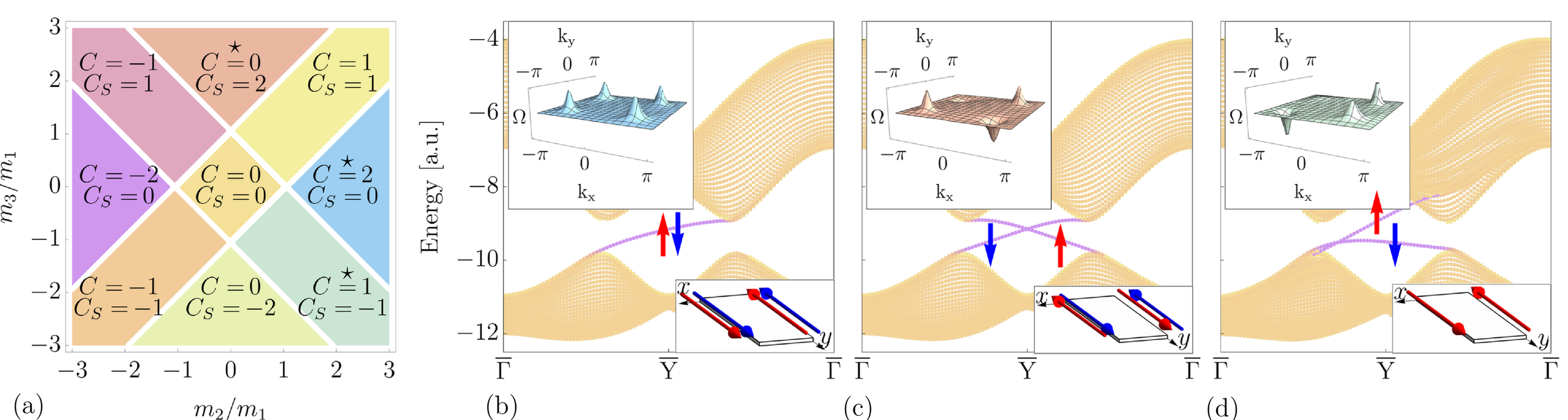}
    \caption{\justifying 
    (a) Phase diagram of the electronic model as a function of the mass terms
    $m(\mathbf{k})= m_1+m_2\sin k_x+m_3\sin k_y$ (with $m_1>0$), which gap out the 2D WPs.
    (b)--(d) 
    Edge band structure of the states
    within the ten outermost unit cells 
    of a 30 cell ribbon with mass terms~\eqref{eq:Massterm}
    for the phase with $C=2$, $m_1=m_3=0$, $m_2=0.5$. (c) Quantum spin Hall state with $C_S=2$, $m_1=m_2=0$, $m_3=0.5$. (d) Quantum Hall state with spin-polarized current with $C=1$, $C_S=-1$, $m_1=0.4$, $m_2=0.5$, $m_3=-0.5$. In the insets we show the Berry curvature distribution and directions of the spin-polarized edge currents for a chemical potential within the bulk gap.}
    \label{fig:el_phases}
\end{figure*}

To observe anomalous spin-polarized currents we gap out the WPs by breaking the $(2_x\mathcal{T}||2_z)$, $(E||\mathcal{M}_{xy})$ and $(E||\mathcal{M}_{x\overline{y}})$ symmetries. Since these symmetries relate sublattice A with C (B with D), we introduce mass terms in the form of staggered potentials and hoppings 
\begin{equation}\label{eq:Massterm}
    M(\mathbf{k})= m(\mathbf{k})\ \text{diag} (+1, +1, -1, -1) ,
\end{equation}
where $m(\mathbf{k})$ has a positive prefactor $+1$ for the A and B sublattices, and a negative  $-1$  for the C and D sublattices. 
In particular, we consider a phase diagram for the three masses $m(\mathbf{k})= m_1+m_2\sin k_x+m_3\sin k_y$. These mass terms do not mix spin up and down allowing for  spin-polarized valleys.
The onsite term $m_1$ can be implemented using a staggered sublattice potential \cite{doi:10.1126/science.1237240} that breaks inversion $(E||2_z)$, whereas the $k$-dependent terms also break the $(2_x\mathcal{T}||E)$ symmetry of the collinear order. Functionally similar terms can be realized in a quasistatic system with laser driving (see Ref.~\cite{PhysRevB.79.081406, PhysRevB.84.235108} and  SM~\cite{sm_note}), where the light intensity is a tuning parameter between distinct phases.

In a system with a chemical potential near the degeneracies, the low-energy physics is governed by the four spin-polarized WPs. 
We construct low-energy models for all four WPs/valleys (see SM~\cite{sm_note}). Each of the models can be described by the effective Hamiltonian
\begin{equation}\label{Effective_LowEn}
    H^{\text{eff}}_{\alpha}(\mathbf{q}) = \epsilon_{\alpha}(\mathbf{q}) \sigma_0 + \sum_{\mathclap{i,j\in\{x,y\}}} q_i (\mathcal{A}_{\alpha})_{ij}\sigma_j,
\end{equation}
where the WP in each of the four quarters of the BZ is labelled by $\alpha \in \{++,--,+-,-+\}$, 
$\mathbf{q}$ is the displacement from the WP, 
$\sigma_i$ are the Pauli matrices, 
and $\mathcal{A}_{\alpha}$ is the coefficient matrix. The symmetries of the spin-wallpaper group relate the four matrices $\mathcal{A}_{\alpha}$ with each other. The term $\epsilon_{\alpha}(\mathbf{q})$ is a linear shift of the bands in energy.
A mass term that opens a gap in Eq.~\eqref{Effective_LowEn} has the form $m_{\alpha}\sigma_z$. 
Each valley contributes to the total (spin) Chern number of the split bands as
\begin{align}
    C^{\alpha} = \pm \tfrac{\text{sign} (m_{\alpha}\det \mathcal{A}_{\alpha})}{2}.\label{Chern_num}
\end{align}
We compute the phase diagram by calculating the Chern number and spin Chern number $C_S = C_{\downarrow} - C_{\uparrow}$ contribution from each of the crossings and summing them up [see Fig.~\ref{fig:el_phases}(a)]. For distinct phases, we show as insets in Fig.~\ref{fig:el_phases}(b)--(d) the Berry curvature distribution~\cite{PhysRevLett.117.052001}. Non-zero values of the (spin) Chern number imply the presence of topological boundary states for a model with edges. 
We compute the band structure for ribbons with termination in the $x$-direction and 30 unit cells. 
In Fig.~\ref{fig:el_phases}(b)--(d) we display the projection of the bands onto ten unit cells at the left edge. In addition, we show the spin character for the emerging edge states. In the phase with the bulk Chern number $C=2$ and $C_S=0$ we observe two boundary states with opposite spins and positive Fermi velocity for the chemical potential in the bulk gap. When $C=0$ and $C_S=2$ the states are spin-polarized and have opposite Fermi velocities. In the phase with $C=1$ and $C_S=-1$ there is a spin-up polarized edge state with positive Fermi velocity inside the bulk gap.

\textit{Electronic transport.---}
The model shows a topological response, as the non-trivial Chern number corresponds to a quantized transverse current.
In the semiclassical limit the Hall current has the form~\cite{PhysRevLett.99.236809,RevModPhys.82.1959}
\begin{equation}\label{semiclas_Hall_current}
    \mathbf{j} = \frac{e^2}{\hbar} \int  \frac{\mathrm{d}^2 \mathbf{k}}{(2\pi)^2} f(\mathbf{k}) \mathbf{E} \times \mathbf{\Omega}(\mathbf{k}),
\end{equation}
where $f(\mathbf{k})$ is the Fermi-Dirac distribution, $\mathbf{E}$ is the electric field, $\mathbf{\Omega}(\mathbf{k})$ is the Berry curvature. We choose the chemical potential $\mu$ within the conduction band, above the gap energy $|m|$, i.e. $|m| < E < \mu$. 
The total current contains contributions from four Fermi pockets, described by the spin-polarized low-energy models.
Integrating the expression for each of the valleys in the limit of a small gap $|m| \ll \mu$ and temperature $T=0$, we obtain an analytical expression for the transverse (spin) conductivity
\begin{align}
    \sigma_{xy} &= \frac{e^2}{2h} \sum_{\alpha} \text{sign} (m_{\alpha} \det \mathcal{A}_{\alpha}) +O(m_{\alpha}/\mu),\\
    \sigma_{xy}^S &= \frac{e}{8\pi} \sum_{\alpha} (-1)^{s(\alpha)}\text{sign} (m_{\alpha} \det \mathcal{A}_{\alpha})+ O(m_{\alpha}/\mu),
\end{align}
where we also consider the
spin current $\mathbf{j}_S = (\hbar/2e)(\mathbf{j}_{\downarrow}-\mathbf{j}_{\uparrow})$. Here, $s(\alpha) = 0$ for the crossings $\alpha$ with spin down, and $s(\alpha) = 1$ for spin up. The main contribution to the conductivity is due to the quantized anomalous current if the gap is small $|m| \ll \mu$. 
For the case when the chemical potential is
inside the gap,
we further compute the temperature dependence of the transverse (spin) conductivity using Eq.~\eqref{semiclas_Hall_current}. In the non-trivial phases, the conductivity reaches quantized values for small temperatures (see SM~\cite{sm_note}).

\begin{figure}
    \centering
    \includegraphics[width=\linewidth]{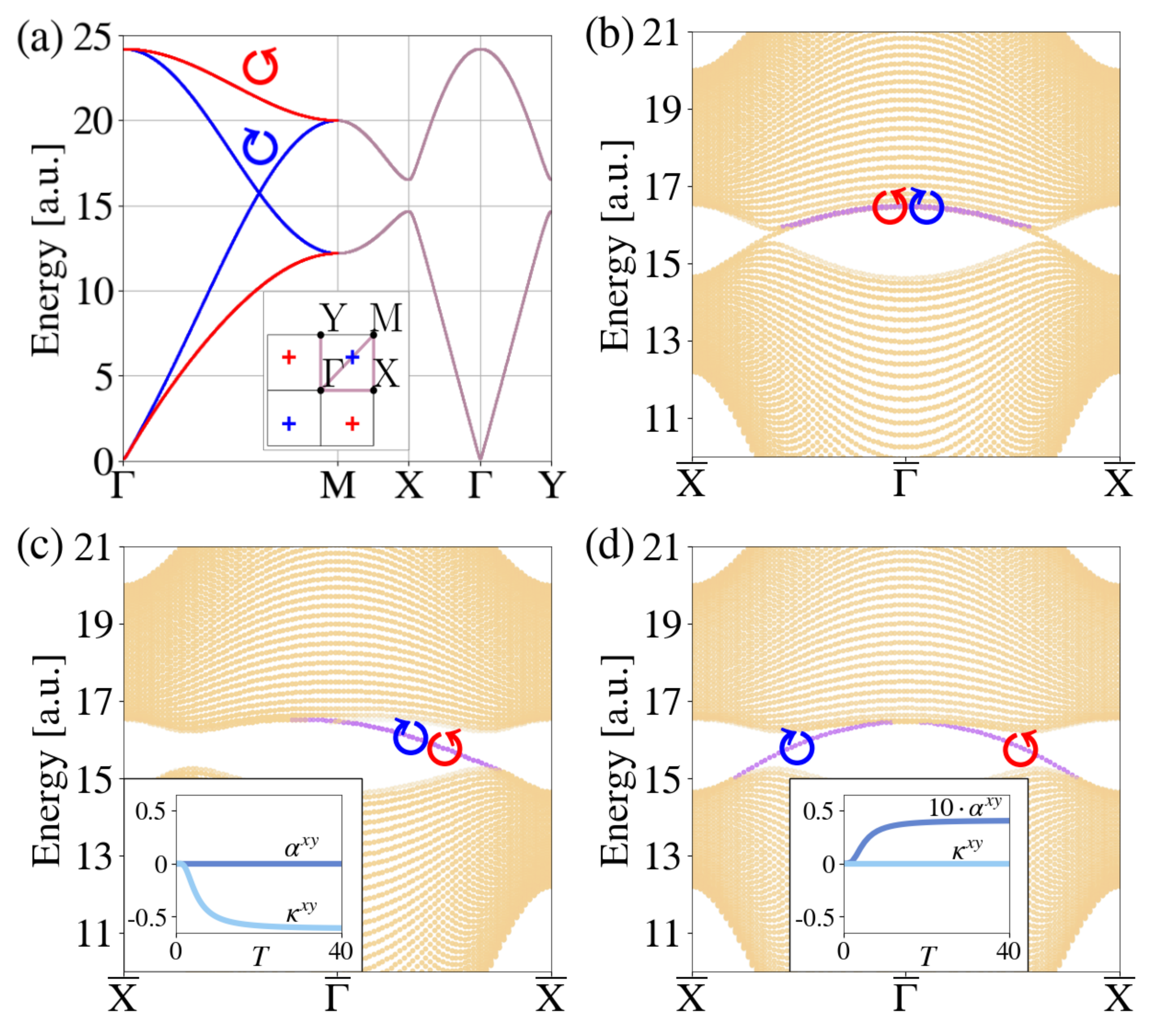}
    \caption{\justifying (a) Magnon band structure of the Hamiltonian~\eqref{eq:LSWH} for the parameters $J=2$, $L=-3$, $L' = L/2$, $L'' = L/5$ with the chirality indicated in red (blue) for right-(left-)handedness.
    The inset shows the BZ with the positions of the WPs. (b)--(d) Edge spectra on a ribbon geometry with 50 unit cells in $y$-direction projected onto ten unit cells at the lower ($y=0$) boundary where the colour indicates the localization on the edge. (c)--(d)  We consider gapped phases with (c) $C=2$, $C_\vartheta = 0$ with mass parameters $m_2 = 0.5$, $m_1 = m_3 = 0$ and (d) $C=0$, $C_\vartheta = 2$ with mass parameters $m_3 = 0.5$, $m_1 = m_2 = 0$. The insets show the transverse thermal conductivity $\kappa^{xy}$ $[k_\mathrm{B}^2/\hbar]$ and the magnon spin Nernst coefficient $\alpha^{xy}$ $[k_\mathrm{B}/\hbar]$ as a function of temperature $T$.}
    \label{fig:MagnonBulkBands}
\end{figure}

\textit{Magnon model.---}
We study the magnon band structure of a minimal model for the altermagnetic wallpaper group p$^{\overline{1}}$4$^{\overline{1}}$m$^1$m for the lattice shown in Fig.~\ref{fig:SWG6}(a). The four sites in the unit cell result in four magnon bands which inherit the altermagnetic spin-wallpaper group symmetries of the exchange interactions. 
The spin interaction Hamiltonian for the localized magnetic moments reads
\begin{align}
    \mathcal{H} = J \sum_{\braket{ij}} \mathbf{S}_i \mathbf{S}_j + L^{\scriptscriptstyle(\scriptstyle\prime\scriptscriptstyle / \scriptstyle\prime\prime\scriptscriptstyle)} \sum_{\braket{\hspace{-.05cm}\braket{ij}\hspace{-.05cm}}} \mathbf{S}_i \mathbf{S}_j,
\end{align}
with $J$ the nearest-neighbor coupling strength between opposite-spin sublattices and $L^{\scriptscriptstyle(\scriptstyle\prime\scriptscriptstyle / \scriptstyle\prime\prime\scriptscriptstyle)}$ the next-nearest-neighbor coupling strengths between same-spin sublattices. 
The couplings are shown in the lattice structure in the SM~\cite{sm_note}.
Performing the standard linear spin-wave theory we obtain the magnon Hamiltonian in momentum space $\mathcal{H}_\mathrm{LSW} = \sum_\mathbf{k} \psi_\mathbf{k}^\dagger H_\mathbf{k}^{\vphantom{\dagger}} \psi_\mathbf{k}^{\vphantom{\dagger}} $ in terms of the bosonic operators $\psi_\mathbf{k} = (a_\mathbf{k}^{\vphantom{\dagger}},b_{-\mathbf{k}}^\dagger,c_\mathbf{k}^{\vphantom{\dagger}},d_{-\mathbf{k}}^\dagger)^{\text{T}}$ with
\begin{align}\label{eq:LSWH}
    H_\mathbf{k} = S \mqty(h_0 & \gamma_1 & \gamma_2 & \gamma_3 \\
                         \gamma_1^* & h_0 & \gamma_3 & \gamma_4 \\
                         \gamma_2^* & \gamma_3^* & h_0 & \gamma_1 \\
                         \gamma_3^* & \gamma_4^* & \gamma_1^* & h_0 ),
\end{align}
where $h_0 = 4J-2L-L'-L''$, $\gamma_1 = 2J\cos\tfrac{k_x}{2}$, $\gamma_2 = 2L\cos\tfrac{-k_x+k_y}{2} + L'e^{i\tfrac{k_x+k_y}{2}} + L''e^{-i\tfrac{k_x+k_y}{2}}$, $\gamma_3 = 2J\cos\tfrac{k_y}{2}$, and $\gamma_4 = 2L\cos\tfrac{k_x+k_y}{2} + L'e^{i\tfrac{-k_x+k_y}{2}} + L''e^{i\tfrac{k_x-k_y}{2}}$. The Hamiltonian is diagonalized with a Bogoliubov transformation \cite{Colpa1978} and the resulting magnon spectrum is shown in Fig.~\ref{fig:MagnonBulkBands}(a). One finds that the magnon bands are pairwise degenerate along the mirror lines,
but are in general split in the rest of the BZ. 
A distinct chirality, corresponding to the direction of precession of the spin around the axis of the collinear magnetic order, can be assigned to the individual magnon bands. Since the collinear order in $z$-direction preserves the $z$-component of the total spin $S^z = \sum_{i} S^z_{i}$ with $i$ running over all sites, we use $\vartheta =\braket{S^z}/\hbar$ to specify the chirality of the bands \cite{Cheng2016Nernst}. In particular, we have $\vartheta=+1$ for right-handed and $\vartheta=-1$ for left-handed magnon modes and we indicate them with curved arrows \textcolor{red}{$\boldsymbol\circlearrowleft$} and \textcolor{blue}{$\boldsymbol\circlearrowright$}, respectively. 
The magnon bands exhibit four symmetry-related WPs formed by same-chirality bands, which are protected by the $(2_x\mathcal{T}||2_z)$ symmetry. For each of the crossings we find a $\pi-$Berry phase. 
Calculating the magnon spectrum 
on a ribbon of finite width in $y$-direction, we find topological edge states between the WPs. In the spectrum shown in Fig.~\ref{fig:MagnonBulkBands}(b), the bands are projected onto the lower ($y = 0$) edge. We find two degenerate bands of opposite chirality per edge.

Next, we study the properties of the gapped system obtained by introducing the mass term of Eq.~\eqref{eq:Massterm} with $m(\mathbf{k})= m_1+m_2\sin k_x +m_3\sin k_y$, which breaks the $(2_x\mathcal{T}||2_z)$, $(E||\mathcal{M}_{xy})$ and $(E||\mathcal{M}_{x\overline{y}})$ symmetries, thereby splitting the WPs. We obtain the phase diagram for the ratio of the mass parameters $m_2/m_1$ and $m_3/m_1$ by calculating the Chern number $C$ of the lower two bands.  
Further, we calculate the magnon analog of the spin Chern number, the chirality Chern number $C_\vartheta = C_{+1} - C_{-1}$ with $C_{\pm1}$ the Chern number for the bands with chirality $\pm 1$. 
The phase diagram we obtain is in full agreement with the one determined for the low-energy electron model shown in Fig.~\ref{fig:el_phases}(a) with $C_\vartheta = C_{S}$. Details on the calculation and the resulting phase diagram for the magnon model are described in the SM~\cite{sm_note}. In the topological phases, i.e., phases with finite (chirality) Chern number, we calculate the spectra on a ribbon of finite width in $y$-direction and show the projection onto the lower ($y=0$) edge in \mbox{Fig.~\ref{fig:MagnonBulkBands}(c)--(d)}. 
For each edge we find two degenerate edge states of opposite chirality in the phase with $C=2$ and $C_\vartheta=0$ [Fig.~\ref{fig:MagnonBulkBands}(c)] and two counter-propagating edge states with opposite chirality for $C=0$ and $C_\vartheta=2$ [Fig.~\ref{fig:MagnonBulkBands}(d)].

\textit{Thermal response.---}
The non-trivial topology of gapped magnon bands can be studied experimentally via thermal transport measurements. 
In the semiclassical limit, in systems with a gap with $C\neq 0$, the finite Berry curvature leads to a thermal Hall current $\mathbf{j} = \kappa^{xy} \hat{z}\times \nabla T$ when applying a longitudinal thermal gradient \cite{Matsumoto2011thermal}. In the case of $C_\vartheta \neq 0$, one finds the magnon spin Nernst effect with the chirality current $\mathbf{j}_\vartheta = \alpha^{xy} \hat{z}\times \nabla T$ as the analogue of the spin Hall effect \cite{Cheng2016Nernst}.
The thermal Hall conductivity $\kappa^{xy}(T)$ and the magnon spin Nerst coefficient $\alpha^{xy}(T)$ for a 2D magnon system are given by \cite{Matsumoto2011thermal,Cheng2016Nernst}
\begin{align}
    \kappa^{xy}(T) &= -\frac{k_\mathrm{B}^2 T}{(2\pi)^2 \hbar} \sum_n \int \limits_\mathrm{BZ} \mathrm{d}^2 \mathbf{k} \, c_2(\varrho_n)\, \Omega_n (\mathbf{k}), \\
    \alpha^{xy}(T) &= -\frac{k_\mathrm{B}}{(2\pi)^2 \hbar} \sum_n \int \limits_\mathrm{BZ} \mathrm{d}^2 \mathbf{k} \, c_1(\varrho_n)\, \vartheta_n \Omega_n (\mathbf{k}),
\end{align}
where $\varrho_n$ is the Bose-Einstein distribution function, $\Omega_n(\mathbf{k})$ the Berry curvature, and $\vartheta_n$ the chirality of band $n$. Further, $c_1(x) = (1+x)\ln (1+x) - x\ln x$,
$c_2(x) = (1+x)(\ln \frac{1+x}{x})^2 - (\ln x)^2 - 2\mathrm{Li}_2(-x)$, and $\mathrm{Li}_2(x)$ is the dilogarithm (see SM \cite{sm_note}). 
The insets in \mbox{Fig.~\ref{fig:MagnonBulkBands}(c)--(d)} show the thermal Hall conductivity and the magnon spin Nernst coefficient as a function of temperature for the discussed topological phases. 
We find a finite thermal Hall effect in the phases with $C\neq 0$ and a finite magnon spin Nernst effect in the phases with $C_\vartheta\neq 0$. In the topologically trivial phase, both responses are zero. The response is not quantized and its strength depends on the parameters of the model and the mass term.

\textit{Conclusion.---} 
In this work we studied point degeneracies between bands of the same spin quantum number in 2D altermagnets. Similar types of degeneracies have recently been theoretically investigated for a mirror symmetric electronic model~\cite{antonenko2024mirror}, where the gapped phase is characterized by a mirror Chern number. Further, the quantum anomalous Hall effect was theoretically predicted in antiferromagnetic CrO~\cite{2023npjCM,10.1063/5.0147450}.
We systematically extend the study of these novel WPs by providing a classification of the degeneracies based on symmetries for the spin-wallpaper groups which provides the basis for an extended search for candidate materials \cite{sodequist2024twodimensional}. 
Additionally to an electronic model, we provide a first example of an altermagnetic magnon model exhibiting WPs involving bands of the same  chirality.
Spin (chirality) polarization of the electronic (magnonic) states in the vicinity of the WPs together with the absence of the total magnetization of the altermagnet allows for unique topological responses. 
Each WP exhibits a parity anomaly~\cite{PhysRevD.29.2366,semenoff_anomaly_PRL,wenbin_parity_anomaly_PRB} leading to topological currents (see SM~\cite{sm_note}) under symmetry-breaking mass terms. The currents give rise to the quantum Hall and quantum spin Hall effect for electrons, and the thermal Hall and spin Nernst effect for magnons.
The mass terms can be controlled, for example, by light interacting with the electrons (magnons), allowing
to drive the 2D altermagnet across different topological 
phase transitions (see SM~\cite{sm_note}). 
In addition, circularly polarized light can induce a valley Hall effect~\cite{PhysRevLett.99.236809},
since right and left circularly polarized light couples differently to electrons from valleys with opposite  Berry curvature. 
Similarly, a magnon valley thermal Hall effect is expected to arise in insulating 2D altermagnets~\cite{PhysRevB.102.075407}.

\textit{Acknowledgments.---} 
We wish to thank Moritz Hirschmann, Niclas Heinsdorf, and Paul McClarty for useful discussions.
This work is funded by the Deutsche Forschungsgemeinschaft (DFG, German Research Foundation) – TRR 360 – 492547816.

\bibliographystyle{apsrev4-1}
\bibliography{bibliography.bib}

\clearpage

\end{document}


\title{Supplemental Material for
\\
``Topological responses from gapped Weyl points in 2D altermagnets"}

\date{\today}

\author{Kirill Parshukov}
\email{k.parshukov@fkf.mpg.de}
\thanks{Shared first author.}
\affiliation{Max Planck Institute for Solid State Research, Heisenbergstrasse 1, D-70569 Stuttgart, Germany}

\author{Raymond Wiedmann}
\email{r.wiedmann@fkf.mpg.de}
\thanks{Shared first author.}
\affiliation{Max Planck Institute for Solid State Research, Heisenbergstrasse 1, D-70569 Stuttgart, Germany}

\author{Andreas P. Schnyder}
\email{a.schnyder@fkf.mpg.de}
\affiliation{Max Planck Institute for Solid State Research, Heisenbergstrasse 1, D-70569 Stuttgart, Germany}

\setcounter{figure}{0}
\makeatletter 
\renewcommand{\thefigure}{S\@arabic\c@figure} 

\setcounter{equation}{0}
\makeatletter 
\renewcommand{\theequation}{S\@arabic\c@equation} 

 \setcounter{table}{0}
\makeatletter 
\renewcommand{\thetable}{S\@Roman\c@table} 

 \setcounter{section}{0}
\makeatletter 
\renewcommand{\thesection}{S\@Roman\c@section} 

\makeatother

\maketitle
\onecolumngrid

\widetext 


\tableofcontents

 \vspace{0.8cm}

This Supplemental Material (SM) is divided into two parts. The first one contains
seven subsections, where we describe the electronic model from the main text, the altermagnetic wallpaper-group classification, and the simplest altermagnetic two-site model with spin-polarized Weyl points. In Sec.~\ref{appendix_symRep} we present the electronic tight-binding model and its symmetries representations. The construction of the low-energy models is discussed in Sec.~\ref{appendix_lowEn}. Using these models we compute anomalous currents in the semiclassical regime (see Sec. \ref{appendix_semiclassical_tr}). The possible realization and tunability of the symmetry-breaking mass terms are discussed in Sec. \ref{appendix_light_ind_HE}. In Sec.~\ref{appendix_valley_HE} we consider the valley Hall effect for a topologically trivial phase due to the interaction with circularly polarized light. The two-site altermagnetic model with spin-polarized Weyl points is discussed in Sec.~\ref{appendix_2sites}. 
Finally, in Sec.~\ref{altermagnetic_WP_appendix} we conclude 
the first part of this SM by presenting the classification of spin-wallpaper groups corresponding to altermagnets.
The second part of this SM contains details on the altermagnetic magnon model. Information on the lattice model and the diagonalizaion of the magnon Hamiltonian is given in Sec.~\ref{sec:MagnonLatticeLSWT}. In Sec.~\ref{sec:MagnonChiralityBerry} we discuss the chirality and Berry curvature of the magnon bands and show the phase diagram of the model under the symmetry-breaking mass terms. In Sec.~\ref{sec:MagnonResponse} we provide details on the thermal response calculations.

\newpage

\section{Electronic Model}
\subsection{Tight-binding model and symmetry representaions}\label{appendix_symRep}
We consider the Wyckoff positions $(\tfrac{1}{4},\tfrac{1}{4}, \uparrow), \ (-\tfrac{1}{4},\tfrac{1}{4}, \downarrow), \ (-\tfrac{1}{4},-\tfrac{1}{4}, \uparrow) \ \text{and} \ (\tfrac{1}{4},-\tfrac{1}{4}, \downarrow)$, denoting them as A, B, C \text{and} D, to construct the tight-binding model. We choose the alignment of the magnetic moments along the $z$-axis so that at A, C (B, D) sites they point in the same direction, corresponding to the arrow in the Wyckoff position. The symmetric model is given by
\begin{align}
    H(\mathbf{k}) &= \begin{pmatrix}
        H_0(\mathbf{k}) + N \Sigma_z&0\\
        0 & H_0(\mathbf{k}) - N \Sigma_z\\
    \end{pmatrix},\\
    H_0(\mathbf{k}) &=  \begin{pmatrix}
        \epsilon&t_{\text{AB}}&t_{\text{AC}}&t_{\text{AD}}\\
        t_{\text{AB}}^* & \epsilon&t_{\text{BC}}&t_{\text{BD}}\\
        t_{\text{AC}}^* & t_{\text{BC}}^*&\epsilon&t_{\text{CD}}\\
        t_{\text{AD}}^* & t_{\text{BD}}^*&t_{\text{CD}}^*&\epsilon\\
    \end{pmatrix},
\end{align}
where we define
\begin{align}
    \epsilon &=0,\  N = 8,\\ 
    t_{\text{AB}}&=-t_{\text{CD}}=4i\sin\tfrac{k_x}{2},\\
    t_{\text{AC}}&=2(\cos\tfrac{k_x-k_y}{2} + \cos\tfrac{k_x+k_y}{2} - i \sin\tfrac{k_x+k_y}{2}),\\
    t_{\text{AD}}&=t_{\text{BC}}=4i \sin\tfrac{k_y}{2}, \\
    t_{\text{BD}}&=2(\cos\tfrac{k_x-k_y}{2} + \cos\tfrac{k_x+k_y}{2} + i \sin\tfrac{k_x-k_y}{2}).
\end{align}
The interaction with the staggered magnetic moments is described by $\Sigma_z = \text{diag} (1, -1, 1, -1)$.
The tight-binding model describes a system with symmetries of the spin-wallpaper group p$^{\overline{1}}$4$^{\overline{1}}$m$^1$m in the cell-periodic convention. The unitary transformation from unit-cell to cell-periodic convention has the form
\begin{align}
    U &= \sigma_0 \otimes \begin{pmatrix}
    e^{i \tfrac{k_x + k_y}{4}}&0&0&0\\
    0&e^{i\tfrac{-k_x+k_y}{4}}&0&0\\
    0&0&e^{i\tfrac{-k_x-k_y}{4}}&0\\
    0&0&0&e^{i\tfrac{k_x-k_y}{4}}\\
    \end{pmatrix} .
\end{align}
The transformation is given by $H(\mathbf{k}) = U^{\dagger} H_{\text{unit-cell}}(\mathbf{k})U$.
The symmetry operations have the following representations in the unit-cell convention
\begin{align}
    D_{(2_x||\mathcal{M}_x)} &= i\sigma_x \otimes \begin{pmatrix}
    0&1&0&0\\
    1&0&0&0\\
    0&0&0&1\\
    0&0&1&0\\
    \end{pmatrix}, \  D_{(2_x||4_z)} = i\sigma_x \otimes 
    \begin{pmatrix}
    0&0&0&1\\
    1&0&0&0\\
    0&1&0&0\\
    0&0&1&0\\
    \end{pmatrix},\\
    D_{(E||2_z)} &= \sigma_0 \otimes \begin{pmatrix}
    0&0&1&0\\
    0&0&0&1\\
    1&0&0&0\\
    0&1&0&0\\
    \end{pmatrix},\ D_{(2_x\mathcal{T}||E)} = \sigma_0 \otimes \begin{pmatrix}
    1&0&0&0\\
    0&1&0&0\\
    0&0&1&0\\
    0&0&0&1\\
    \end{pmatrix} \mathcal{K}.
\end{align}
\subsection{Low-energy model construction}\label{appendix_lowEn}
To construct the low-energy electronic models at the band crossings we expand the Bloch Hamiltonian in the vicinity of the crossings up to first order. 
There are no matrix elements that couple spin up and down degrees of freedom, the crossings are between bands with the same spin and, therefore, it is enough to consider each spin block of the tight-binding model separately. The obtained $4 \times 4$ matrix for the spin $\downarrow$ crossing takes the form $H_{\downarrow}^{\text{lin}}(\mathbf{k}) = q_x \Sigma_x+q_y\Sigma_y +\Sigma_0$, where $(k_x^0, k_y^0)$ is the momentum-space location of the crossing point, $ \Sigma_i$ are $4 \times 4$ matrices, and $q_x = (k_x-k_{x}^0), q_y = (k_y-k_{y}^0)$ are the displacements. The $\Sigma_0$ matrix has two degenerate eigenstates $\psi_1$ and $\psi_2$, corresponding to the band crossing. We project the $4 \times 4$ matrix onto the subspace spanned by $\psi_1$ and $\psi_2$ introducing an effective Hamiltonian
\begin{align}
    P &= |\psi_1\rangle\langle\psi_1|+|\psi_2\rangle\langle\psi_2|,\\
    H_{\downarrow}^{\text{eff}}(\mathbf{k}) &= PH_{\downarrow}^{\text{lin}}P.
\end{align}
In the same way we obtain the effective mass term
\begin{align}
    M^{\text{eff}} &= PM(k_x^0,k_y^0)P.
\end{align}
Both the effective Hamiltonain and the effective mass term are linear combinations of the Pauli $\sigma$ matrices in the restricted subspace. We change the basis with a unitary transformation that diagonalizes the effective mass term. In the new basis, the mass term is the $\sigma_z$ matrix, while the Hamiltonian is a linear combination of $\sigma_0$, $\sigma_x$, and $\sigma_y$. We list the obtained effective Hamiltonians for the four crossings. For our tight-binding model the crossing points are located at $\mathbf{k}^0 = (\pm1.91, \pm1.91)$ and we obtain
\begin{subequations} \label{eq_low_ene_eff_hams}
\begin{align}
    H_{++}^{\text{eff}}&=\epsilon_{++}(\mathbf{q}) \sigma_0 + q_{\alpha}\sigma_x +q_{\beta}\sigma_y,\\
    H_{--}^{\text{eff}}&=\epsilon_{--}(\mathbf{q}) \sigma_0 -q_{\alpha}\sigma_x +q_{\beta}\sigma_y,\\
     H_{-+}^{\text{eff}}&=\epsilon_{-+}(\mathbf{q}) \sigma_0 + q_{\gamma}\sigma_x +q_{\delta}\sigma_y,\\
    H_{+-}^{\text{eff}}&=\epsilon_{+-}(\mathbf{q}) \sigma_0 -q_{\gamma}\sigma_x +q_{\delta}\sigma_y,
\end{align}
\end{subequations}
where we introduce $q_{\alpha} = 1.0q_x+1.4q_y, q_{\beta}=1.2q_x-0.7q_y, q_{\gamma} = 1.2(q_x-q_y), q_{\delta} = -q_x-q_y$, $\epsilon_{++}(\mathbf{q})=-9.3 - 0.3 q_x - 0.3 q_y$, $\epsilon_{--}(\mathbf{q})=-9.3 + 0.3 q_x + 0.3 q_y$, $\epsilon_{-+}(\mathbf{q})=-9.3 + 0.3 q_x - 0.3 q_y$,
and
$\epsilon_{+-}(\mathbf{q})=-9.3 - 0.3 q_x + 0.3 q_y$. For each of the models, the mass term is $-0.9\sigma_z$. With this we introduce
the four velocity matrices $\mathcal{A}_{\alpha}$, allowing us to write
the effective Hamiltonians as $H^{\textrm{eff}}_{\alpha}(\mathbf{q}) - \epsilon_{\alpha}(\mathbf{q}) \sigma_0 = [(\mathcal{A}_{\alpha})_{11}q_x+(\mathcal{A}_{\alpha})_{21}q_y]\sigma_x+[(\mathcal{A}_{\alpha})_{12}q_x+(\mathcal{A}_{\alpha})_{22}q_y]\sigma_y$. The determinant of the velocity matrix has opposite signs at the crossings that are related by the two-fold rotation, i.e.,  $\text{sign } \det \mathcal{A}_{++} = -\text{sign } \det \mathcal{A}_{--}=\text{sign } \det \mathcal{A}_{+-} = -\text{sign } \det \mathcal{A}_{-+}$.

\subsection{Semiclassical transport}\label{appendix_semiclassical_tr}
To derive the transverse currents within the semiclassical regime we consider a generic $2 \times 2$ effective Hamiltonian given by
 $H = d_0(\mathbf{q})\sigma_0 + d_1(\mathbf{q})\sigma_x + d_2(\mathbf{q})\sigma_y +m \sigma_z$.
Note that all of our
 derived low-energy models~\eqref{eq_low_ene_eff_hams} are exactly 
 of this form. 
The conduction band of this $2 \times 2$ Hamiltonian  is described by the eigenenergy and eigenstate 
\begin{align}
    E_{+}(\mathbf{q}) &= d_0 + \sqrt{d_1^2 + d_2^2 + m^2},\\
    |+,\mathbf{q}\rangle  &= \frac{1}{\sqrt{2d(d-m)}}
    \begin{pmatrix}
    m+d \\ d_1+i d_2
    \end{pmatrix},\\
    d &= \sqrt{d_1^2 + d_2^2 + m^2}.
\end{align}
For the conduction band, the Berry connection and Berry curvature are given by \cite{PhysRevB.74.085308}
\begin{align}
    A_{+i}(\mathbf{q}) &= \langle +,\mathbf{q}|i \partial_{q_i} |+,\mathbf{q}\rangle = \frac{d_1\partial_{q_i}d_2 -d_2\partial_{q_i} d_1}{2d(d+m)},\\
    \mathbf{\Omega}_{+}(\mathbf{q}) &= \frac{m}{2d^3} (\partial_{q_x} d_1 \partial_{q_y} d_2 - \partial_{q_y} d_1 \partial_{q_x} d_2) \mathbf{e_z}.
\end{align}
Writing the Berry curvature using the velocity matrix
\begin{align}
    \mathcal{A}&=\begin{pmatrix}
    \partial_{q_x} d_1 & \partial_{q_y} d_1 \\ 
    \partial_{q_x} d_2 & \partial_{q_y} d_2
    \end{pmatrix},
\end{align}
we obtain
\begin{align}
    \mathbf{\Omega}_{+}(\mathbf{q}) &= \frac{m}{2d^3} \det \mathcal{A} \ \mathbf{e_z},
\end{align}
and with this expression one can determine the valley contribution to the band Chern number $C = \tfrac{\text{sign } (m\det \mathcal{A})}{2}$. To calculate the Hall current we use the semiclassical expression~\cite{PhysRevLett.99.236809,RevModPhys.82.1959}
\begin{equation}
    \mathbf{j} = \frac{e^2}{\hbar} \int \frac{\text{d}^2 \mathbf{k}}{(2\pi)^2} f(\mathbf{k}) \, \mathbf{E} \times \mathbf{\Omega}_+(\mathbf{k}),
\end{equation}
where $f(\mathbf{k})$ is the Fermi-Dirac distribution and $\mathbf{E}$ is an applied electric field.
For the chemical potential in the conduction band, i.e., $|m| < E_+ < \mu$, the Fermi surface has four pockets enclosing the Weyl points and the states in each of the pockets are spin-polarized. We compute the currents individually for each crossing at $T=0$ as
\begin{align}
    \mathbf{j}  &= \frac{e^2}{\hbar} \int \limits_{\mathclap{|m| < E_+(q, \phi) < \mu}} \frac{q\text{d}q\text{d}\phi}{(2\pi)^2}\frac{ m\det \mathcal{A}}{2(d_1^2(\phi, q)+d_2^2(\phi, q)+m^2)^{3/2}}\mathbf{E} \times \mathbf{e_z},\\
    d_0(q, \phi) &\equiv q\overline{\mathcal{A}_0}(\phi),\\
    d_1(q, \phi) &= q(\mathcal{A}_{11}\cos\phi+\mathcal{A}_{12}\sin \phi) \equiv q \overline{\mathcal{A}_1}(\phi),\\
    d_2(q, \phi) &= q(\mathcal{A}_{21}\cos\phi+\mathcal{A}_{22}\sin \phi) \equiv q \overline{\mathcal{A}_2}(\phi),
\end{align}
where we inserted the Berry curvature, used the polar coordinates $q_x = q \cos{\phi}, q_y = q \sin{\phi}$, and defined functions $\overline{\mathcal{A}_i}(\phi)$. First, we integrate over $q$ with the limits $0 < q^2 < q^2_{F}(\phi)$, where $q^2_{F}(\phi)$ is determined by the expression $\mu = \sqrt{d_1^2(q_{F},\phi)+ d_2^2(q_{F},\phi)+ m^2} + d_0(q_{F},\phi)$. In our case $d_0^2(q,\phi) < d_1^2(q,\phi), d_2^2(q,\phi)$ for finite $q$ and $|m|\ll \mu$ we have
\begin{equation}
    q_F(\phi) = O \left( \frac{\mu}{ 
\sqrt{\overline{\mathcal{A}_1}^2(\phi) +\overline{\mathcal{A}_2}^2(\phi)} - \overline{\mathcal{A}_0}(\phi) 
} \right).
\end{equation}
Integrating the current we obtain
\begin{align}
\mathbf{j}  &= -\frac{e^2 m \det \mathcal{A}}{2\hbar} \int \limits_{0}^{2\pi} \frac{\text{d}\phi}{(2\pi)^2} \frac{1}{\overline{\mathcal{A}_1}^2(\phi) +\overline{\mathcal{A}_2}^2(\phi)}   \left(\frac{ 1}{\sqrt{q_{F}^2(\overline{\mathcal{A}_1}^2(\phi) +\overline{\mathcal{A}_2}^2(\phi)) +m^2}} - \frac{ 1}{|m|}\right) \mathbf{E} \times \mathbf{e_z} \\
&= -\frac{e^2 m \det \mathcal{A}}{2\hbar} \int \limits_{0}^{2\pi} \frac{\text{d}\phi}{(2\pi)^2} \frac{1}{\overline{\mathcal{A}_1}^2(\phi) +\overline{\mathcal{A}_2}^2(\phi)} \frac{ 1}{\sqrt{q_{F}^2(\overline{\mathcal{A}_1}^2(\phi) +\overline{\mathcal{A}_2}^2(\phi)) +m^2}}
\mathbf{E} \times \mathbf{e_z}\\
&+\frac{e^2 m\det \mathcal{A}}{2h} \frac{1}{|m\det\mathcal{A}|} \mathbf{E} \times \mathbf{e_z}\\
&= \left(\frac{e^2 \text{sign } (m \det \mathcal{A})}{2h} - O\left(\frac{m}{\mu}\right)\right)\mathbf{E} \times \mathbf{e_z}.
\end{align}
The first term in the final expression is the quantized contribution. It is the anomalous current under the parity breaking. The quantized term contributes the most if the gap is much smaller than the chemical potential $|m| \ll \mu$.
Summing up the contributions of the four Fermi pockets we obtain the charge and spin currents $\mathbf{j}$ and  \mbox{$\mathbf{j}^S = (\hbar/2e)(\mathbf{j}_{\downarrow}-\mathbf{j}_{\uparrow})$} as
\begin{align}
    \mathbf{j} &= \frac{e^2}{2h} \Bigl(\text{sign } (m_{++} \det \mathcal{A}_{++})+\text{sign } (m_{--} \det \mathcal{A}_{-}) \Bigr. +\text{sign } (m_{+-} \det \mathcal{A}_{+-})
    +\text{sign } (m_{-+} \det \mathcal{A}_{-+})\\
    &+\Bigl.O(\frac{m}{\mu})\Bigr)
    \mathbf{E} \times \mathbf{e_z},\\
    \mathbf{j}^S &= \frac{e}{8\pi} \Bigl(\text{sign } (m_{++} \det \mathcal{A}_{++})+\text{sign } (m_{--} \det \mathcal{A}_{-}) \Bigr. -\text{sign } (m_{+-} \det \mathcal{A}_{+-})
    -\text{sign } (m_{-+} \det \mathcal{A}_{-+})\\
    &+\Bigl.O(\frac{m}{\mu})\Bigr)
    \mathbf{E} \times \mathbf{e_z} .
\end{align} 

\begin{figure}[h!]
    \centering
    \includegraphics[width=\textwidth]{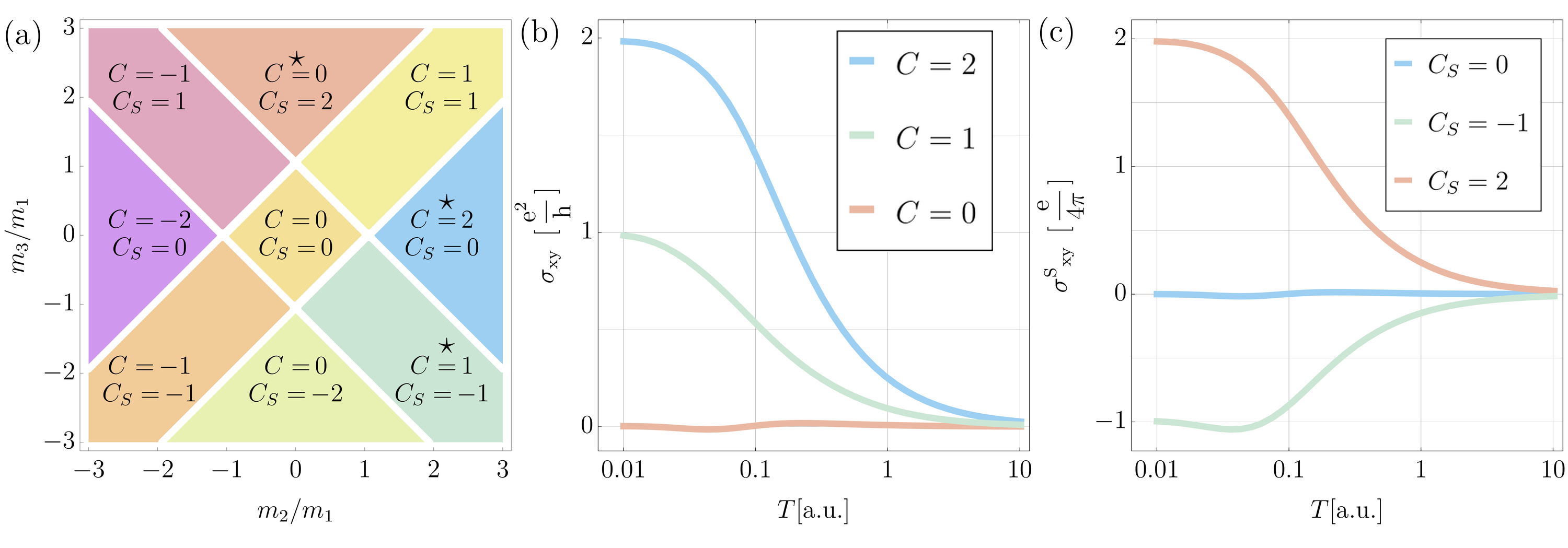}
   \caption{
   (a) Phase diagram for the electronic model as a function of the mass terms
    $m(\mathbf{k})= m_1+m_2\sin k_x+m_3\sin k_y$ (with $m_1>0$), which gap out the 2D WPs.
    (b), (c) Transverse charge and spin conductivity, respectively, as a function of temperature $T$ with the chemical potential inside the gap. 
    }
    \label{fig:phase_conductivity}
\end{figure}

We numerically compute the temperature dependence of the transverse conductivities and plot them in Fig.~\ref{fig:phase_conductivity} for the three phases discussed in the main text. The quantized values at low temperatures are in agreement with the obtained formulas.

\subsection{Light-induced Hall effect}\label{appendix_light_ind_HE}

One of the possible ways to gap out the WPs and get a non-trivial response is to apply circularly polarized light~\cite{PhysRevB.79.081406, PhysRevB.84.235108}. Neglecting the magnetic components, a circularly polarized light field of frequency $\omega$ is described by the vector potential
\begin{equation}
    \mathbf{A}(t) = \mathbf{e}_x A_x  \cos \omega t + \mathbf{e}_y A_y\sin \omega t. 
\end{equation}
Using the Peierls substitution in the case of space-independent fields we get the time-dependent Hamiltonian $H(\mathbf{k},t)=H(\mathbf{k}-e\mathbf{A}(t))$.
We write the Hamiltonian in the Floquet Brillouin zone for $T=2\pi/\omega$
\begin{equation}
    H_n = \frac{1}{T}\int \limits^T_0 \text{d}t e^{in\omega t}H(\mathbf{k},t).
\end{equation}
The $t$-independent terms do not contribute to the expression above. The only relevant terms are given by $e(A_x \mathcal{A}_{11} \cos \omega t + A_y\mathcal{A}_{12}\sin \omega t)\sigma_x+e(A_x \mathcal{A}_{21} \cos \omega t + A_y\mathcal{A}_{22}\sin \omega t)\sigma_y$. Under the Fourier transform for $n=\pm1$ we get
\begin{align}
    H_{1} &=\frac{e}{2}(A_x \mathcal{A}_{11} +i A_y\mathcal{A}_{12})\sigma_x+\frac{e}{2}(A_x \mathcal{A}_{21}  +i A_y\mathcal{A}_{22})\sigma_y,\\
    H_{-1} &=\frac{e}{2}(A_x \mathcal{A}_{11} -i A_y\mathcal{A}_{12})\sigma_x+\frac{e}{2}(A_x \mathcal{A}_{21}  -i A_y\mathcal{A}_{22})\sigma_y.
\end{align}
For large frequencies the $H_{\pm1}$ can be treated as a perturbation to the initial Hamiltonian (in the new notation $H_0$), i.e., we have
\begin{align}
    H_{\text{eff}}^{\text{Floquet}} &= H_0 + \frac{1}{\omega} [H_{-1},H_1] =H_0 + \frac{e^2}{4\omega}(A_x \mathcal{A}_{11} -i A_y\mathcal{A}_{12})(A_x \mathcal{A}_{21}  +i A_y\mathcal{A}_{22}) [\sigma_x, \sigma_y]\\
    &+\frac{e^2}{4\omega}(A_x \mathcal{A}_{21}  -i A_y\mathcal{A}_{22})(A_x \mathcal{A}_{11} +i A_y\mathcal{A}_{12})[\sigma_y,\sigma_x]\\
    &=H_0 - \frac{e^2A_x A_y \det \mathcal{A}}{\omega}\sigma_z.
\end{align}
In this picture, we obtain a mass term that opens a gap. The mass term depends on $\det \mathcal{A}$ making the Chern number contribution independent of an effective low-energy model $C = \mp \tfrac{\text{sign } (A_x A_y)}{2}$. The effect is similar to the mass term introduced in the main text as $\sin(k_x)\, \text{diag} (+1, +1, -1, -1)$. 

\begin{figure}[h!]
    \centering
    \includegraphics[width=.3\linewidth]{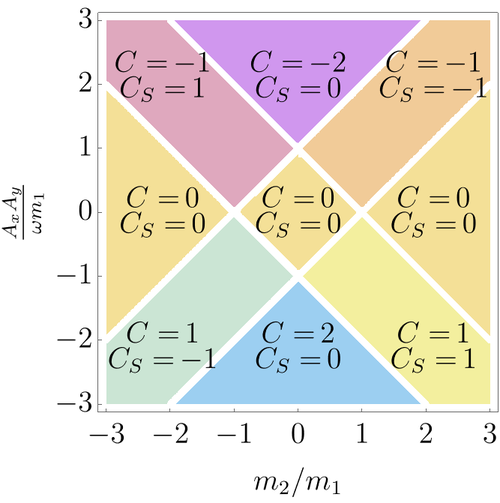}
   \caption{
    Phase diagram for the electronic model as a function of the mass term ratios
    $\tfrac{m_2}{m_1}$ and $\tfrac{e^2A_x A_y}{\omega m_1}$ (with $m_1>0$), which gap out the 2D WPs.
  }
    \label{fig:phase_Light}
\end{figure}

In addition we can consider other symmetry breaking terms due to the interaction with a substrate. For the model there are two possible onsite mass terms
\begin{align}
m_1 \left(
\begin{array}{cccc}
 1 & 0 & 0 & 0 \\
 0 & 1 & 0 & 0 \\
 0 & 0 & -1 & 0 \\
 0 & 0 & 0 & -1 \\
\end{array}
\right)+m_2 \left(
\begin{array}{cccc}
 1 & 0 & 0 & 0 \\
 0 & -1 & 0 & 0 \\
 0 & 0 & -1 & 0 \\
 0 & 0 & 0 & 1 \\
\end{array}
\right).
\end{align}
At the four crossings the effective mass terms have the form 
\begin{align}
    m_{++} &= m_{--}=(-m_1+m_2)\sigma_z,\\
    m_{+-} &= m_{-+} =(-m_1-m_2)\sigma_z.
\end{align}
For our $\mathcal{A}_{\alpha}$ matrices the total Chern numbers are given by
\begin{align}
    C_{++} &=  \text{sign } (-m_1+m_2 - \frac{e^2A_x A_y}{\omega}), \\
    C_{+-} &= \text{sign } (-m_1-m_2- \frac{e^2A_x A_y}{\omega}),\\
    C_{--} &=  \text{sign } (m_1-m_2 - \frac{e^2A_x A_y }{\omega}), \\
    C_{-+} &= \text{sign } (m_1+m_2- \frac{e^2A_x A_y }{\omega}). 
\end{align}
We show the phase diagram for a fixed $m_1>0$ in Fig.~\ref{fig:phase_Light}. The ratio $m_2/m_1$ is fixed by the substrate, however, $\frac{e^2A_x A_y}{\omega m_1}$ can be varied by the light intensity. This tunability allows for a number of gapped phases with $C\in\{-2,-1,0,1,2\}$ and $C_S\in\{-1,0,1\}$.
\subsection{Valley Hall effect}\label{appendix_valley_HE}
In the topologically trivial phase with $C=C_S=0$ both spin and charge transverse currents are compensated. However, we expect topological transport phenomena in a system interacting with circularly polarized light. 
Circularly polarized light couples differently to electrons from different valleys.
Thus, the number of electrons that are excited into the conduction band differs for each valley,
giving rise to a valley Hall effect~\cite{PhysRevB.77.235406,PhysRevLett.99.236809,PhysRevLett.108.196802}.

To study this effect, we consider interband transitions due to the interaction with circularly polarized light and derive the selection rules. The interband matrix element of the canonical momentum operator is given by
\begin{align}
    P_i(\mathbf{q}) &= m_e \langle u_c|\tfrac{\partial H}{\partial q_i}|u_v\rangle,\label{circ_pol_light}
\end{align}
where $m_e$ is the free electron mass, $|u_v\rangle$ is the valence band state, $|u_c\rangle$ is the conduction band state.
A circularly polarized optical field with polarization $\pm$ couples to electrons with the coupling strength 
\begin{align}
    P_{\pm}(\mathbf{q}) &= P_x(\mathbf{q}) \pm i P_y(\mathbf{q}).\label{circ_pol_light}
\end{align}
For a generic Dirac Hamiltonian $H(\mathbf{q}) = (\mathcal{A}_{11}q_x+\mathcal{A}_{21}q_y)\sigma_x+(\mathcal{A}_{12}q_x+\mathcal{A}_{22}q_y)\sigma_z+m\sigma_y$ the probability for transitions from the valence
to the conduction band at the WP crossing is given by
\begin{align}
    P_{\pm} &= (\mathcal{A}_{12}\pm \mathcal{A}_{21}\tfrac{m}{|m|})^2 + (\mathcal{A}_{22} \mp \mathcal{A}_{11}\tfrac{m}{|m|})^2 \\
    &= \sum \mathcal{A}_{ij}^2 \mp 2 \det \mathcal{A} \ \text{sign } m ,
\end{align}
which depends on the determinant of the velocity matrix
$\mathcal{A}$.
Hence, the transition probability is in general
enhanced or suppressed depending on the sign
of $\det \mathcal{A}$, i.e., the sign of the Berry curvature.
%
%
This creates a carrier imbalance between valleys
with opposite Berry curvature. 
Since the Hall conductivity is proportional to the density of photoinduced electrons (or holes) times 
the Berry curvature,
the carrier imbalance leads to 
an anomalous Hall conductivity, namely,
to the valley Hall effect~\cite{PhysRevB.77.235406}.

\subsection{Two-site model}\label{appendix_2sites}

The model described in the main text is not the simplest electronic model showing spin-polarized WPs protected by $(2_x\mathcal{T}||2_z)$ symmetry. 
In fact there is a simpler model with only two sites in the unit cell,
which has the symmetries of the spin-wallpaper group p$^1$2$^{\overline{1}}$g$^{\overline{1}}$g (related to the wallpaper group \#8, p2gg).
The altermagnetic spin-wallpaper group    p$^1$2$^{\overline{1}}$g$^{\overline{1}}$g contains the symmetries
\begin{align}
    b^{\infty} &= \text{SO(2)} \rtimes \{(E||E), (2_x\mathcal{T}||E)\},\\
    h &= \{(E||E),(E||2_z)\},\\
    G_{NS} &=  h + (2_x||\mathcal{M}_x(\tfrac{1}{2},\tfrac{1}{2})) h,\\
    G &= b^{\infty} \times G_{NS},
\end{align}
where $b^{\infty}$ is the spin-only part of the group, corresponding to symmetries of the collinear order. All symmetry operations have the form $(s||g)$, where $s$ acts on the spin degree of freedom and $g$ acts on the lattice. 
The symmetry group of the lattice sites is denoted by $h$. The sites with localized magnetic moments are related by the non-symmorphic operation $(2_x||\mathcal{M}_x(\tfrac{1}{2},\tfrac{1}{2}))$. $G_{NS}$ is the non-trivial part of the spin-wallpaper group, as it combines operations on both spin and lattice degrees of freedom. The total spin-wallpaper group is given by the group $G$.
The $(2_x\mathcal{T}||2_z)$ symmetry quantizes the Berry phase to $\pi$, protecting point degeneracies (i.e., WPs), between bands with the same spin character. 
In systems with the spin-wallpaper group p$^1$2$^{\overline{1}}$g$^{\overline{1}}$g, WPs occur
always in multiples of $4n$  with $n\in\mathbb{N}_0$,
where quartets of WPs are related by the symmetries.

\begin{figure}
    \centering
    \includegraphics[width=0.3\textwidth]{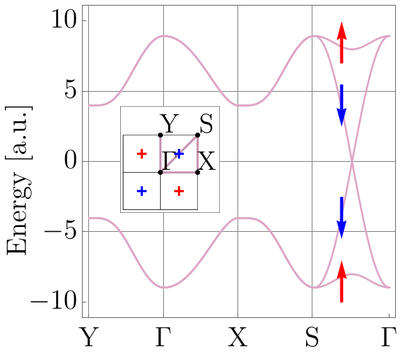}
    \caption{Electronic bands of the altermagnetic two-site tight-binding model, Eq.~\eqref{two_site_model_ham}. The BZ is shown in the inset.}
    \label{fig:SWG8}
\end{figure}

To construct the tight-binding model, we consider the Wyckoff positions $(0,0, \uparrow), (\tfrac{1}{2},\tfrac{1}{2},\downarrow)$. The simplest symmetric model is given by
\begin{subequations} \label{two_site_model_ham}
\begin{align}
    H_{\uparrow}(\mathbf{k}) &= 
    \left(
    \begin{array}{cc}
     -4. \cos k_y & -8 \cos \frac{k_x-k_y}{2} \\
     -8 \cos \frac{k_x-k_y}{2} & 4. \cos k_y \\
    \end{array}
    \right),\\
    H_{\downarrow}(\mathbf{k}) &= 
    \left(
    \begin{array}{cc}
     4. \cos k_y & -8 \cos \frac{k_x+k_y}{2} \\
     -8 \cos \frac{k_x+k_y}{2} & -4. \cos k_y \\
    \end{array}
    \right),\\
    H(\mathbf{k}) &= \begin{pmatrix}
        H_{\downarrow}(\mathbf{k}) &0\\
        0 & H_{\uparrow}(\mathbf{k})\\
    \end{pmatrix}. 
\end{align}
\end{subequations}
This model describes a system with four symmetry-related crossings between the second and third band (see Fig.~\ref{fig:SWG8}). 
The model has nodal line degeneracies between bands with opposite spin along the mirror lines $\Gamma$X, $\Gamma$Y, SX, and SY. The degeneracies are enforced by the symmetries $(2_x||\mathcal{M}_x(\tfrac{1}{2},\tfrac{1}{2})), (\mathcal{T}||\mathcal{M}_x(\tfrac{1}{2},\tfrac{1}{2}))$, and the SO(2) spin rotational symmetry. This degeneracy is lifted at a general position in the BZ, except at the
four points $(\pm \tfrac{\pi}{2}, \pm \tfrac{\pi}{2})$,
where two bands with the same spin from WP crossings
(see Fig.~\ref{fig:SWG8}).

Let us now consider the WP crossing at the position $(\tfrac{\pi}{2}, \tfrac{\pi}{2})$ and construct its low-energy model
\begin{equation}\label{low_en_model}
    H_{\downarrow}(\mathbf{q}) = -q_y\sigma_z + (q_x +q_y)\sigma_x,
\end{equation}
where $q_x$ and $q_y$ are the displacements from the crossing point. This low-energy model describes the crossing between bands with the same spin $\downarrow$. The mass term that opens a gap has the form $m\sigma_y$. This WP (i.e., valley) gives a contribution to the Chern number of the conduction and valence bands given by
\begin{align}
    C = \pm \tfrac{1}{2} \text{sign} (\det \mathcal{A}) \text{sign} (m),
\end{align}
where $\mathcal{A}$ is the matrix of the velocities $H(\mathbf{q}) = q_i \mathcal{A}_{ij}\sigma_j$. 
Importantly, the symmetries of the spin-wallpaper group 
p$^1$2$^{\overline{1}}$g$^{\overline{1}}$g relate these matrices at the four WP crossings, and thus $\det \mathcal{A}$ is the same for all four WPs/valleys. Therefore, the total Chern number for the four crossings depends only on the mass term.

We consider mass terms that preserve the SO(2) symmetry, and therefore do not mix spin up and down. Any onsite terms or hoppings between sites 
of the same sublattice do not break the $(2_x\mathcal{T}||2_z)$ symmetry, and therefore do not lift the degeneracy. 
Hence, the mass term must involve hoppings between sites 
of different sublattices. Here we consider general hoppings of this type between the nearest neighbor sites
\begin{align}
    m(\mathbf{k}) &= t_1 e^{\frac{1}{2} i (k_x+k_y)}+t_2 e^{-\frac{1}{2} i (k_x+k_y)}
    +t_3 e^{\frac{1}{2} i (k_x-k_y)}+t_4 e^{\frac{1}{2} i (-k_x+k_y)},\\
    M_{\downarrow}(\mathbf{k}) &= M_{\uparrow}(\mathbf{k}) = \begin{pmatrix}
        0&m(\mathbf{k})\\
        m^*(\mathbf{k})&0
    \end{pmatrix}.
\end{align}
To observe the quantum Hall effect with $C=\pm2$ we choose $t_1=t_2=t_3=t_4=-i$. With this choice the mass term in momentum space has the form $4 \sigma_y\cos\tfrac{k_x}{2} \cos\tfrac{k_y}{2}$. Despite the $k$-dependence of the mass term, it has the same sign for all four crossings. 
For the quantum-spin Hall state with $C_S = \pm2, \ C = 0$ we take $t_1=t_2=-t_3=-t_4=i$,  yielding $m (\mathbf{k} ) = 4 \sigma_y\sin\tfrac{k_x}{2} \sin\tfrac{k_y}{2}$. With this choice the sign of $m (\mathbf{k} )$ is alternating between quarters of the Brillouin zone. Crossings with different spins have opposite signs of the mass term. The discussed mass terms can arise due to spin-orbit coupling that breaks the altermagnetic symmetries.

Another possibility is to break the $(E||2_z)$ symmetry with the real hoppings $t_1 = -t_2 = t_3 = -t_4 = -1$, giving the mass term
$m (\mathbf{k} ) = 4\sigma_y \cos\tfrac{k_y}{2} \sin\tfrac{k_x}{2}$. In this case, $C = 0$ and $ C_S = 0$, so the phase is trivial. However, one still can observe the valley Hall effect. 
\begin{figure}
    \centering
\includegraphics[width=0.3\textwidth]{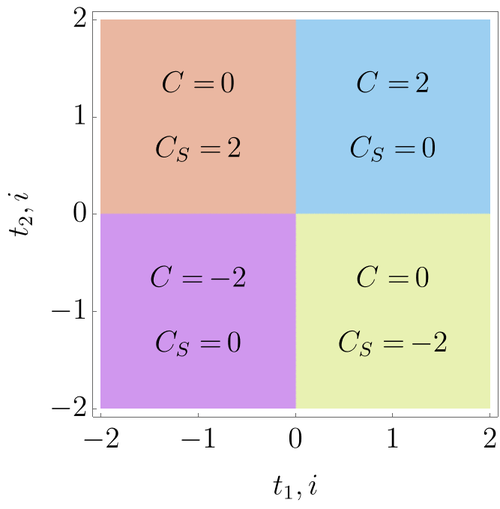}
    \caption{Phase diagram for imaginary hoppings $t_1 = t_2$, $t_3=t_4$ in the two-site model.} 
    \label{fig:Phase_diag_2site}
\end{figure}

We compute the transverse Hall current that originates from states near the WP at $(\tfrac{\pi}{2},\tfrac{\pi}{2})$. As before, we use the semiclassical expression~\cite{PhysRevLett.99.236809,RevModPhys.82.1959}
\begin{equation}\label{trans_current}
    \mathbf{j}_{\downarrow} = \frac{e^2}{\hbar} \int \frac{\text{d}^2 q}{(2\pi)^2} f(\mathbf{q}) \, \mathbf{E} \times \mathbf{\Omega}_{+,\downarrow}(\mathbf{q}),
\end{equation}
where $f(\mathbf{q})$ is the Fermi-Dirac distribution and $\mathbf{E}$ is the electric field. The current is fully spin-polarized. We choose the chemical potential $\mu$ within the conduction band, above the gap energy $|m|$, i.e. $|m| < E_+ < \mu$. The energy and the state for the upper band have the form
\begin{align}
    E_{+, \downarrow}(\mathbf{q}) &= \sqrt{q_y^2 + (q_x + q_y)^2 + m^2},\\
    |+,\downarrow,\mathbf{q}\rangle  &= \frac{1}{\sqrt{N_+}}
    \begin{pmatrix}
    -q_y + \sqrt{q_y^2 + (q_x + q_y)^2 + m^2} \\ (q_x + q_y + i m)
    \end{pmatrix},\\
    N_{+,\downarrow}(\mathbf{q}) &= 2E_+(\mathbf{q})(E_+(\mathbf{q})-q_y).
\end{align}
For the state $|+,\downarrow,\mathbf{q}\rangle$ we compute the Berry connection and curvature
\begin{align}
    A_{++,\downarrow}^x(\mathbf{q}) &= \langle +,\downarrow,\mathbf{q}|i \partial_{q_x} | +,\downarrow,\mathbf{q}\rangle = \frac{m}{2E_+(\mathbf{q})(E_+(\mathbf{q})-q_y)},\\
    A_{++,\downarrow}^y(\mathbf{q}) &= \langle  +,\downarrow,\mathbf{q}|i \partial_{q_y} | +,\downarrow,\mathbf{q}\rangle = \frac{m}{2E_+(\mathbf{q})(E_+(\mathbf{q})-q_y)},\\
    \Omega_{+,\downarrow}^z(\mathbf{q}) &= \partial_{qx}A_{++,\downarrow}^y(\mathbf{q}) - \partial_{qy}A_{++,\downarrow}^x(\mathbf{q}) = -\frac{m}{2E_+(\mathbf{q})^3}.
\end{align}
We note that the minus sign is in full agreement with the fact that the velocity matrix for the WP at $(\tfrac{\pi}{2},\tfrac{\pi}{2})$ has negative determinant, i.e., $\det \mathcal{A} = -1$.
The expression (\ref{trans_current}) for the chemical potential in the conduction band, i.e. $|m| < E_+ < \mu$, takes the form 
\begin{align}
    \mathbf{j}_{\downarrow}  &= \frac{e^2}{\hbar} \int \limits_{\mathclap{|m| < E_+(q, \phi) < \mu}} \frac{q\text{d}q\text{d}\phi}{(2\pi)^2} \frac{-m}{2(m^2 + q^2 + q^2\sin{2\phi} +q^2\sin^2{\phi})^{3/2}}\mathbf{E} \times \mathbf{e_z},
\end{align}
where we inserted the Berry curvature and used the polar coordinates $q_x = q \cos{\phi}, q_y = q \sin{\phi}$. We integrate over $q$ with the limits $0 < q^2 < \tfrac{\mu^2 - m^2}{1 + \sin{2\phi} + \sin^2{\phi}}$ which yields
\begin{align}
    \mathbf{j}_{\downarrow} &= \frac{e^2}{\hbar} \int \limits_0^{2\pi} \frac{\text{d}\phi}{2(2\pi)^2}  \frac{m - \mu \ \text{sign }m }{\mu (1 + \sin{2\phi} + \sin^2{\phi})}\mathbf{E} \times \mathbf{e_z} \\
    &= - \frac{e^2}{2h}   \left( \text{sign }m - \frac{m}{\mu }\right)\mathbf{E} \times \mathbf{e_z}.
\end{align}
For a small energy gap, the main contribution to the current is due to the parity anomaly~\cite{PhysRevD.29.2366,semenoff_anomaly_PRL}, and the direction is determined by the sign of the mass term, and consequently the Chern number.

\subsection{Altermagnetic wallpaper groups}\label{altermagnetic_WP_appendix}
We construct spin-wallpaper groups following Ref.\ \cite{jiang2023enumeration} by first identifying a normal subgroup $h$ of a wallpaper group, which corresponds to sublattice symmetries. The quotient group then relates the sublattices. 
We restrict ourselves to the altermagnetic case without a supercell, where the symmetry relating sublattices is not a $2_z$ rotation or translation. The obtained wallpaper groups are listed in Tab.~\ref{tab:AM_SWG}. We use the common notation for the wallpaper groups, modifying them with the superscript showing the action of the spin degree of freedom, where $1$ is the trivial operation and $\overline{1}$ is the spin flip. 
For the wallpaper groups \#11 (p4mm) and \#12 (p4gm) there are three altermagnetic space groups with different normal subgroups, i.e., sublattice symmetry groups. 
In the main text, we list the groups with symmetries that protect point degeneracies. The minimal multiplicity of point degeneracies is four. In groups \#6.1, \#7.1, \#8.1, and \#9.1 the point degeneracy is located at a general position and related with three other points by mirror and rotational symmetries. In models with symmetries of groups \#10.1, \#11.1, \#11.2, \#11.3, \#12.1, \#12.2 and \#12.3, two-fold point degeneracies are related by the four-fold rotational symmetry. In group \#17.1 the number of crossings is increased due to the 6-fold rotational symmetry. We also note the altermagnetic splitting. A general degeneracy between bands with opposite spins is the line degeneracy. There is always spin degeneracy at the $\Gamma$ point in altermagnets due to the symmetry relating the sublattices. The number of nodal lines passing the $\Gamma$ point is always even due to SO(2) and time-reversal symmetry $(2_x\mathcal{T}||E)$, which relates points in the BZ with the same spin character.
\begin{table}[h]
\renewcommand{\arraystretch}{1.4} %
\centering
\begin{tabular}{ |c|c| }
 \hline
 \# & notation\\
 \hline
3.1 & p$^{\overline{1}}$m\\
4.1 &p$^{\overline{1}}$g\\
5.1 &c$^{\overline{1}}$m\\
6.1 &p$^1$2$^{\overline{1}}$m$^{\overline{1}}$m\\
7.1 &p$^1$2$^{\overline{1}}$m$^{\overline{1}}$g\\
8.1 &p$^1$2$^{\overline{1}}$g$^{\overline{1}}$g\\
9.1 &c$^1$2$^{\overline{1}}$m$^{\overline{1}}$m\\
10.1 &p$^{\overline{1}}$4\\
&\\
\hline
\end{tabular}
\begin{tabular}{ |c|c| }
 \hline
 \# & notation\\
 \hline
11.1 &p$^{\overline{1}}$4$^{\overline{1}}$m$^1$m\\
11.2 &p$^{\overline{1}}$4$^1$m$^{\overline{1}}$m\\
11.3 &p$^1$4$^{\overline{1}}$m$^{\overline{1}}$m\\
12.1 &p$^{\overline{1}}$4$^{\overline{1}}$g$^1$m\\
12.2 &p$^{\overline{1}}$4$^1$g$^{\overline{1}}$m\\
12.3 &p$^1$4$^{\overline{1}}$g$^{\overline{1}}$m\\
14.1 &p$^1$3$^{\overline{1}}$m1\\
15.1 &p$^1$31$^{\overline{1}}$m\\
17.1 &p$^1$6$^{\overline{1}}$m$^{\overline{1}}$m\\
\hline
\end{tabular}
\caption{Altermagnetic wallpaper groups.}
\label{tab:AM_SWG}
\end{table}

\newpage

\section{Magnon Model}

\subsection{Lattice model and linear spin-wave theory}\label{sec:MagnonLatticeLSWT}

\begin{figure}
    \centering
    \includegraphics[width=.6\linewidth]{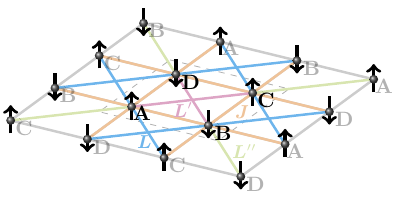}
    \caption{Lattice and coupling parameters for the magnon model for the altermagnetic wallpaper group p$^{\overline{1}}$4$^{\overline{1}}$m$^1$m with $J$ the nearest-neighbor coupling strength between opposite-spin sublattices and $L^{\scriptscriptstyle(\scriptstyle\prime\scriptscriptstyle / \scriptstyle\prime\prime\scriptscriptstyle)}$ the next-nearest-neighbor coupling strengths between same-spin sublattices. Lines of the same color have the same coupling parameter. The unit cell is marked by a dashed lines. Note that only couplings involving at least one site of the marked unit cell are shown.}
    \label{fig:magnoncouplings}
\end{figure}

We consider the altermagnetic wallpaper group p$^{\overline{1}}$4$^{\overline{1}}$m$^1$m and the lattice and the spin coupling parameters shown in Fig. \ref{fig:magnoncouplings}. The spin interaction Hamiltonian for the localized magnetic moments reads
\begin{align}
    \mathcal{H} = J \sum_{\braket{ij}} \mathbf{S}_i \mathbf{S}_j + L^{\scriptscriptstyle(\scriptstyle\prime\scriptscriptstyle / \scriptstyle\prime\prime\scriptscriptstyle)} \sum_{\braket{\hspace{-.05cm}\braket{ij}\hspace{-.05cm}}} \mathbf{S}_i \mathbf{S}_j,
\end{align}
with $J$ the nearest-neighbor coupling strength between opposite-spin sublattices and $L^{\scriptscriptstyle(\scriptstyle\prime\scriptscriptstyle / \scriptstyle\prime\prime\scriptscriptstyle)}$ the next-nearest-neighbor coupling strengths between same-spin sublattices. Note that, in particular, the spatial anisotropy of the ferromagnetic next-nearest neighbor couplings breaks the space-time inversion and time-inversion with fractional translation symmetries found in antiferromagnets and, therefore, reflects the altermagnetic symmetries of the system and causes the chirality-splitting of the bands. 

We rewrite the spin Hamiltonian in terms of bosonic creation and annihilation operators with the linearized Holstein-Primakoff transformation \cite{HolsteinPrimakoff1940} for
\begin{align*}
    \text{sublattice A (and C with $c^{\vphantom{\dagger}}, c^\dagger$):}&\quad\qquad S^x_\mathrm{A} = \tfrac{\sqrt{2S}}{2}(a^{\vphantom{\dagger}} + a^\dagger), \quad\qquad S^y_\mathrm{A} = \tfrac{\sqrt{2S}}{2i}(a^{\vphantom{\dagger}} - a^\dagger), \quad\qquad S^z_\mathrm{A} = S - a^\dagger a^{\vphantom{\dagger}}, \\
    \text{sublattice B (and D with $d^{\vphantom{\dagger}}, d^\dagger$):}&\quad\qquad S^x_\mathrm{B} = \tfrac{\sqrt{2S}}{2}(b^{\vphantom{\dagger}} + b^\dagger), \quad\qquad S^y_\mathrm{B} = -\tfrac{\sqrt{2S}}{2i}(b^{\vphantom{\dagger}} - b^\dagger), \quad\qquad S^z_\mathrm{B} = -(S - b^\dagger b^{\vphantom{\dagger}}),
\end{align*}
 with $S$ the spin magnitude. The linearization is valid in the case of $S \gg \braket{n}$, i.e., for a small number of magnon excitations, because then terms beyond quadratic order in the creation/annihilation operators (i.e., magnon-magnon interactions) can be neglected. After performing a Fourier transformation, we obtain the bilinear magnon Hamiltonian in momentum space $\mathcal{H}_\mathrm{LSW} = \sum_\mathbf{k} \psi_\mathbf{k}^\dagger H_\mathbf{k}^{\vphantom{\dagger}} \psi_\mathbf{k}^{\vphantom{\dagger}} $ with \mbox{$\psi_\mathbf{k} = (a_\mathbf{k}^{\vphantom{\dagger}},b_{-\mathbf{k}}^\dagger,c_\mathbf{k}^{\vphantom{\dagger}},d_{-\mathbf{k}}^\dagger)^{\text{T}}$} and
\begin{align}
    H_\mathbf{k} =S \mqty(h_0 & \gamma_1 & \gamma_2 & \gamma_3 \\
                         \gamma_1^* & h_0 & \gamma_3 & \gamma_4 \\
                         \gamma_2^* & \gamma_3^* & h_0 & \gamma_1 \\
                         \gamma_3^* & \gamma_4^* & \gamma_1^* & h_0 ) \; ,
\end{align}
where 
\begin{align*}
    h_0 &= 4J-2L-L'-L''\; , \\
    \gamma_1 &= 2J\cos\tfrac{k_x}{2}\; ,  \qquad \qquad \qquad \;
    \gamma_2 = 2L\cos\tfrac{-k_x+k_y}{2} + L'e^{i\tfrac{k_x+k_y}{2}} + L''e^{-i\tfrac{k_x+k_y}{2}} \; , \\
    \gamma_3 &= 2J\cos\tfrac{k_y}{2} \; , \qquad \qquad \qquad \;
    \gamma_4 = 2L\cos\tfrac{k_x+k_y}{2} + L'e^{i\tfrac{-k_x+k_y}{2}} + L''e^{i\tfrac{k_x-k_y}{2}} \; .
\end{align*}

This Hamiltonian is diagonalized with a Bogoliubov transformation $\phi_\mathbf{k} = T \psi_\mathbf{k}$  in terms of the magnon operators for the modes $\alpha$, $\beta$, $\gamma$, and $\delta$ with \mbox{$\phi_\mathbf{k} = (\alpha_\mathbf{k}^{\vphantom{\dagger}},\beta_{\mathbf{k}}^\dagger,\gamma_\mathbf{k}^{\vphantom{\dagger}},\delta_{\mathbf{k}}^\dagger)^{\text{T}}$}. To preserve the bosonic commutation relations, $T$ must be chosen such that $T\Sigma_z T^\dagger =\Sigma_z$, where $\Sigma_z = \mathrm{diag}(+1, -1, +1, -1)$, i.e., $T$ is a paraunitary matrix \cite{Colpa1978}.

\begin{figure}
    \centering
    \includegraphics[width= \linewidth]{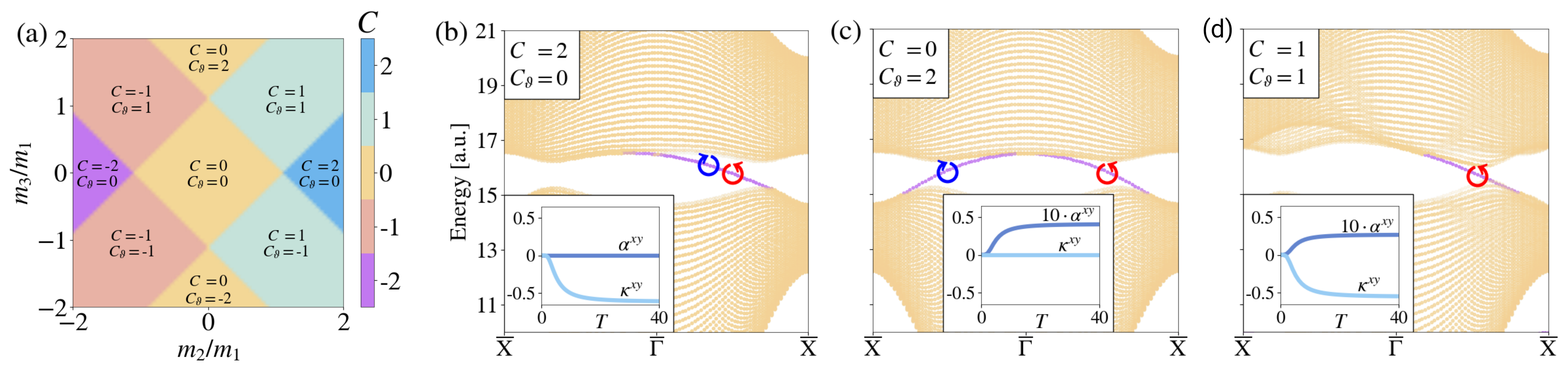}
    \caption{
    (a) Phase diagram of the magnon model as a function of the mass terms
    $m(\mathbf{k})= m_1+m_2\sin k_x+m_3\sin k_y$ (with $m_1>0$), obtained by calculating the Chern number $C$ and chirality Chern number $C_\vartheta$.
    (b)--(d) Magnon edge spectra calculated on a ribbon geometry with 50 unit cells in $y$-direction for different mass terms for the parameters $J=2$, $L=-3$, $L' = L/2$, $L'' = L/5$. The bands are projected onto ten unit cells at the lower ($y=0$) boundary of the ribbon  where the colour indicates the localization on the edge. The magnon chirality of the topological edge states is indicated with curved arrows. The insets show the transverse thermal conductivity $\kappa^{xy}$ in units $[k_\mathrm{B}^2/\hbar]$ and the magnon spin Nernst coefficient $\alpha^{xy}$ in units $[k_\mathrm{B}/\hbar]$ as a function of temperature $T$ for the respective phases. We consider phases with (a) $C=2$, $C_\vartheta = 0$ with mass parameters $m_2 = 0.5$, $m_1 = m_3 = 0$, (b) $C=0$, $C_\vartheta = 2$ with mass parameters $m_3 = 0.5$, $m_1 = m_2 = 0$, and (c) $C=1$, $C_\vartheta = 1$ with mass parameters $m_1 = -0.5$, $m_2 = 0.8$, $m_3 = 0.5$.}
    \label{fig:MagnonEdgeStates}
\end{figure}

\subsection{Magnon chirality, Berry curvature, and phase diagram}\label{sec:MagnonChiralityBerry}
We can assign a distinct chirality to the individual magnon bands, corresponding to the direction of precession of the spin around the axis of the collinear magnetic order. The collinear order in $z$-direction preserves the $z$-component of the total spin $S^z = \sum_{i} S^z_{i}$ with $i$ running over all sites. After the Holstein-Primakoff and the Fourier transformation this yields $S^z = \sum_\mathbf{k} (-a^\dagger_\mathbf{k}a^{\vphantom{\dagger}}_\mathbf{k} + b^\dagger_\mathbf{k}b^{\vphantom{\dagger}}_\mathbf{k} - c^\dagger_\mathbf{k}c^{\vphantom{\dagger}}_\mathbf{k} + d^\dagger_\mathbf{k}d^{\vphantom{\dagger}}_\mathbf{k})$ and performing the Bogoliubov transformation we obtain
\begin{equation}
    S^z = \sum_\mathbf{k} (-\alpha^\dagger_\mathbf{k}\alpha^{\vphantom{\dagger}}_\mathbf{k} + \beta^\dagger_{\mathbf{k}}\beta^{\vphantom{\dagger}}_{\mathbf{k}} - \gamma^\dagger_\mathbf{k}\gamma^{\vphantom{\dagger}}_\mathbf{k} + \delta^\dagger_{\mathbf{k}}\delta^{\vphantom{\dagger}}_{\mathbf{k}})\; , 
\end{equation}
for the $z$-component of the total spin. We use $\vartheta =\braket{0|\zeta S^z \zeta^\dagger|0}/\hbar$ for \mbox{$\zeta \in \{\alpha_\mathbf{k}, \beta_{\mathbf{k}}, \gamma_\mathbf{k}, \delta_{\mathbf{k}}\}$} with $\ket{0}$ the magnon vacuum to specify the chirality of the bands \cite{Cheng2016Nernst}. In particular, we have $\vartheta=+1$ for right-handed and $\vartheta=-1$ for left-handed magnon modes. 
Due to the collinear order of the system, the magnon chirality is conserved and we can use it to decouple the right- and left-handed magnon sectors and calculate Berry curvature, Chern numbers, and topological responses for each magnon sector separately.

The Berry connection for magnon band $n$ is given by \cite{annurev-conmatphys-031620-104715}
\begin{equation}
    \mathcal{A}_n^\nu = i [\Sigma_z T^\dagger \Sigma_z  \partial_{k_\nu} T]_{nn},
\end{equation}
and the out-of-plane component of the Berry curvature is defined as 
\begin{equation}\label{eq:MagnonBerry}
    \Omega_n = \partial_{k_x} \mathcal{A}_n^y - \partial_{k_y} \mathcal{A}_n^x  \; .
\end{equation}
The corresponding Chern number is given by
\begin{equation}
    C_n = \frac{1}{2\pi} \int\limits_\mathrm{BZ} \Omega_n\,  \mathrm{d}\mathbf{k} \; .
\end{equation}

By introducing a mass term, as described in the main text, the system obtains a gap between the lower two and upper two bands. The Chern number for the lower two bands is determined by adding the individual Chern numbers for the right- and left-handed band as $C = C_{+1} + C_{-1}$. Note that $C_{\pm1}$ is in general a sum over the bands below the gap with chirality $\vartheta = \pm 1$. We obtain the analogue of the spin Chern number, the chirality Chern number, by taking the difference between the Chern numbers of the bands of different chirality as $C_\vartheta = C_{+1} - C_{-1}$.
With this, the phase diagram for the magnon model is determined by calculating the Chern and chirality Chern number for different ratios of the mass terms and is shown in Fig.~\ref{fig:MagnonEdgeStates} for $m_1 > 0$, where the colours of the phases correspond to the value of the Chern number. The phase diagram is in full agreement with the one obtained for the electronic model shown in Fig.~\ref{fig:phase_conductivity}(a).

\begin{figure}
    \centering
    \includegraphics[width = \linewidth]{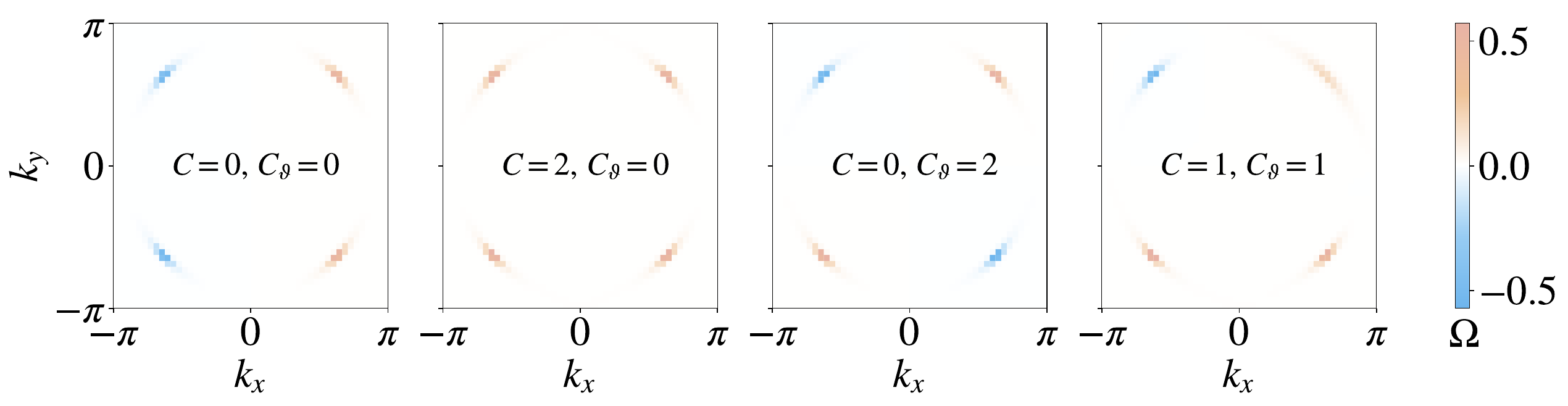}
    \caption{Berry curvature distribution in the BZ of the two bands below the gap for the magnon model with gapped WPs for four distinct phases with different (chirality) Chern numbers. The mass parameters $(m_1, m_2, m_3)$ are, from left to right, $(0.1, 0.0, 0.0)$, $(0.0, 0.1, 0.0)$, $(0.0, 0.0, 0.1)$, and $(0.1, 0.1, 0.1)$.}
    \label{fig:MagnonModelBerryCurv}
\end{figure}

The distribution of the Berry curvature in the BZ for the topologically trivial phase and the three distinct topological phases is shown in Fig.~\ref{fig:MagnonModelBerryCurv}. As expected, we find that the Berry curvature peaks at the positions of the gapped WPs and changes its sign corresponding to the change of the sign of the mass terms in the BZ.

\subsection{Details on thermal response calculations}\label{sec:MagnonResponse}

In the semiclassical limit, in systems with a gap with $C\neq 0$, the finite Berry curvature leads to a transverse thermal Hall current $\mathbf{j} = \kappa^{xy} \hat{z}\times \nabla T$ when applying a longitudinal thermal gradient $\nabla T$. In the case of a finite chirality Chern number $C_\vartheta \neq 0$, one finds the magnon spin Nernst effect with the chirality current $\mathbf{j}_\vartheta = \alpha^{xy} \hat{z}\times \nabla T$ as the analogue of the spin Hall effect in electronic systems.
The thermal Hall conductivity $\kappa^{xy}(T)$ \cite{Matsumoto2011thermal} and the magnon spin Nerst coefficient $\alpha^{xy}(T)$ \cite{Cheng2016Nernst} for a 2D magnon system are given by
\begin{align}
    \kappa^{xy}(T) &= -\frac{k_\mathrm{B}^2 T}{(2\pi)^2 \hbar} \sum_n \int\limits_\mathrm{BZ} \mathrm{d}^2 k \, c_2(\varrho_n)\, \Omega_n (\mathbf{k}), \\
    \alpha^{xy}(T) &= -\frac{k_\mathrm{B}}{(2\pi)^2 \hbar} \sum_n \int\limits_\mathrm{BZ} \mathrm{d}^2 k \, c_1(\varrho_n)\, \vartheta_n \Omega_n (\mathbf{k}),
\end{align}
where $\varrho_n = [ e^{\varepsilon_n(\mathbf{k})/(k_\mathrm{B}T)} -1]^{-1}$ is the bosonic distribution function for energy band $\varepsilon_n(\mathbf{k})$, $\Omega_n(\mathbf{k})$ is the Berry curvature of band $n$ as defined in Eq.~(\ref{eq:MagnonBerry}), and $\vartheta_n$ is the magnon chirality. 
The functions $c_1$ and $c_2$ of the bosonic distribution function are defined as  
\begin{align*}
    c_1(x) &= (1+x)\ln (1+x) - x\ln x \; , \\
    c_2(x) &= (1+x)\left(\ln \frac{1+x}{x}\right)^2 - (\ln x)^2 - 2\mathrm{Li}_2(-x) \; , 
\end{align*}
where $\mathrm{Li}_2(x) = -\int\limits_0^x \frac{\log(1-t)}{t} \mathrm{d}t$ is the dilogarithm. 

The responses for three distinct topological phases are shown as insets in Fig.~\ref{fig:MagnonEdgeStates}(b)--(d) with the corresponding ribbon spectra. The response is not quantized, vanishes at $T=0$, and saturates at high temperatures. Further, the strength of the response depends on gap size (determined by the mass parameters) and the coupling parameters of the model.

\bibliographystyle{apsrev4-1}
\bibliography{bibliography.bib}